\documentclass[a4paper,11pt]{article}
\pdfoutput=1 

\usepackage{jheppub} 
\usepackage{amsmath,amssymb,amsthm,amscd,graphicx}
\usepackage{psfrag}
\usepackage[english]{babel}
\usepackage{float}\input epsf.sty
\usepackage{slashed}
\usepackage[compat=1.1.0]{tikz-feynman}
\usepackage{simpler-wick}

\usepackage{color}

\theoremstyle{definition}

\newcommand{\CL}{{\cal L}}

\newcommand{\CO}{{\cal O}}

\def\bPhi{\boldsymbol{\Phi}}

\newcommand{\re}{{\rm e}}
\newcommand{\ri}{{\rm i}}
\newcommand{\rd}{{\rm d}}

\newcommand{\mZ}{\mathsf{Z}}

\newcommand{\be}{\begin{equation}}
\newcommand{\ee}{\end{equation}}
\newcommand{\ba}{\begin{aligned}}
\newcommand{\ea}{\end{aligned}}
\newcommand{\ben}{\begin{eqnarray}\displaystyle}
\newcommand{\een}{\end{eqnarray}}

\newcommand{\sectiono}[1]{\section{#1}\setcounter{equation}{0}}

\newdimen\tableauside\tableauside=1.0ex
\newdimen\tableaurule\tableaurule=0.4pt
\newdimen\tableaustep
\def\phantomhrule#1{\hbox{\vbox to0pt{\hrule height\tableaurule width#1\vss}}}
\def\phantomvrule#1{\vbox{\hbox to0pt{\vrule width\tableaurule height#1\hss}}}
\def\sqr{\vbox{%
  \phantomhrule\tableaustep
  \hbox{\phantomvrule\tableaustep\kern\tableaustep\phantomvrule\tableaustep}%
  \hbox{\vbox{\phantomhrule\tableauside}\kern-\tableaurule}}}
\def\squares#1{\hbox{\count0=#1\noindent\loop\sqr
  \advance\count0 by-1 \ifnum\count0>0\repeat}}
\def\tableau#1{\vcenter{\offinterlineskip
  \tableaustep=\tableauside\advance\tableaustep by-\tableaurule
  \kern\normallineskip\hbox
    {\kern\normallineskip\vbox
      {\gettableau#1 0 }%
     \kern\normallineskip\kern\tableaurule}%
  \kern\normallineskip\kern\tableaurule}}
\def\gettableau#1{\ifnum#1=0\let\next=\null\else
\squares{#1}\let\next=\gettableau\fi\next}

\newcommand{\bss}{\boldsymbol{\sigma}}
\newcommand{\figref}[1]{Fig.~\protect\ref{#1}}

\title{ \Huge{Trans-series from condensates in the non-linear sigma model}}

\author[a]{Yizhuang Liu}
\author[b]{and Marcos Mari\~no}
\affiliation[a]{Institute of Theoretical Physics,
Jagiellonian University, 30-348 Kraków, Poland}

\affiliation[b]{D\'epartement de Physique Th\'eorique et Section de Math\'ematiques\\
Universit\'e de Gen\`eve, Gen\`eve, CH-1211 Switzerland}

\emailAdd{yizhuang.liu@uj.edu.pl}
\emailAdd{Marcos.Marino@unige.ch}

\abstract{In this work we provide a massless perturbative framework for the
two dimensional non-linear sigma model (NLSM), that allows the computation of the perturbative
series attached to the operator condensates in the operator product expansion (OPE). It is based on
a limit of the quartic linear sigma model (LSM) and is manifestly $O(N)$ symmetric. We show,
at next-to-leading order in the $1/N$ expansion, how this framework reproduces the perturbative
contribution to the two-point function, as well as its first exponentially small correction due to the
condensate of the Lagrangian operator, in full agreement with the exact non-perturbative large $N$ solution.
We also show that, in the full LSM, the physics at the natural UV cutoff indeed decouples from the
NLSM in the IR, in the weak-coupling limit. In particular, we show that the perturbative framework for
the LSM at the cutoff scale is connected to the one in the NLSM. The structure of power divergences
in the LSM regularization also reveals that the first renormalon on the positive Borel axis of the NLSM perturbative
self-energy is an UV renormalon, which cancels against the ambiguity in the condensate.}

\begin{document}
\maketitle
\flushbottom

%
%

\sectiono{Introduction}

The non-linear sigma model (NLSM) \cite{polyakov,bzj} is
one of the most studied quantum field theories (QFTs). It has applications in condensed matter physics and has also served as a theoretical laboratory for general ideas and techniques in QFT. It was found by Polyakov in \cite{polyakov} that the model
is asymptotically free in two dimensions, and it has been often studied as a toy model for the gluon sector of QCD.

In asymptotically free theories, perturbative series are afflicted with infrared (IR) renormalon
singularities which indicate the existence of
non-perturbative corrections \cite{parisi-renormalons} (see \cite{beneke} for a review and a list of references).
In some cases, these corrections can be computed by using the OPE and assuming that the operators appearing there have non-trivial vacuum expectation values, or condensates. This is the method underlying the SVZ sum rules \cite{svz}. In QCD these phenomena are difficult to
study quantitatively, but the two-dimensional NLSM gives a framework
to test these ideas in detail. For example, the NLSM can be studied with an exact large $N$ solution. In this solution, physical observables are obtained as series in $1/N$ in which each coefficient is a well-defined function of the renormalized coupling constant. The asymptotic expansion of these functions at weak coupling reproduces conventional perturbation theory, but they encode in addition non-perturbative effects which can
be studied in detail. Because of this, the large $N$ expansion in the NLSM has been a fascinating laboratory to study the interplay between
perturbative and non-perturbative physics.

One important lesson of early studies of the NLSM is that the $1/N$ expansion has
the structure required to match the OPE picture of \cite{svz},
but with the important proviso that condensates of operators are ambiguous due to
IR renormalons \cite{david2}. The results of the exact $1/N$ expansion have been
explicitly compared to perturbation theory with condensates in some situations \cite{svz-pr}, but this comparison is
plagued with technical difficulties. For example, it was found in \cite{beneke-braun} that the $1/N$ correction to the two-point function in the NLSM can be written as a trans-series, involving an infinite number of non-perturbative corrections, but none of these corrections have been explicitly reproduced with condensate/OPE calculations. The main difficulty is well-known: the field of the NLSM is constrained to live on a sphere. Solving this
constraint explicitly leads to a theory with an infinite number of vertices, in
which the $O(N)$ invariance is not manifest. This makes the proof
of renormalizability of the model already challenging in this representation \cite{blgzj}.
Studying condensates in this approach is even harder,
since condensates respect $O(N)$ invariance. A number of tricks to overcome this problem have been proposed in the past, in particular in \cite{svz-pr} and more recently in \cite{sss}, but explicit calculations remain involved.

In this paper we address these issues with the concrete goal of verifying the first
non-perturbative correction of \cite{beneke-braun} by using
condensate calculus. This correction is due to the Lagrangian operator of the NLSM
and can be regarded as a toy model for the gluon condensate in QCD. The successful comparison
that we make in this paper is a precision test of the idea that non-perturbative
corrections can be reproduced by combining the OPE with non-trivial vevs for the
operators of the theory. A similar test was recently performed for the two-point function
in the Gross--Neveu model \cite{tsc}, which is technically simpler. In order to do the
computation in the NLSM in a feasible way, we appeal to the old idea that the NLSM
can be obtained from the linear sigma model (LSM) with a quartic
potential and a negative squared mass \cite{bzj,bessiszj}, by sending the negative
squared mass to infinity. We find that
this idea can be implemented in detail and leads to
relatively simple expressions for the two-point function of the NLSM, both in the
perturbative and the non-perturbative sector, at least when
working in the $1/N$ expansion. This formalism has the advantage of
preserving $O(N)$ invariance, which makes it technically simpler and also extremely
convenient for incorporating condensates.

A somewhat surprising conclusion of our analysis of
the perturbative expansion of the self-energy is that the first renormalon appearing in the
positive real axis in fact an UV one, as we show in detail by a standard analysis
of momentum regions. This UV renormalon is
cancelled by the ambiguity in the condensate of the Lagrangian operator. Although this
seems to conflict with the standard picture that condensates cancel IR renormalons,
we show that this is not the case: this unusual cancellation is due to the presence of power divergences, and
is in fact the net result of two more conventional cancellations.
The fact that power divergences might lead to UV renormalons in the positive real axis was first
pointed out in \cite{bb-polemass} in the case of the self-energy of a heavy quark. The situation here is
different though, since the UV renormalon in \cite{bb-polemass} is only cancelled when the
heavy quark effective theory is embedded in QCD. A situation more
similar to ours was observed in \cite{jamin-ramon}, in the QCD correlation function of flavor-changing vector currents: at the perturbative level, its longitudinal part suffers from an UV renormalon on the positive real axis, which is also inherited from a quadratic divergence. In the OPE, it is eventually cancelled by the ambiguity of the first operator condensate, like here.

The representation of the NLSM as a limit of the LSM can be implemented in two different ways.
 One possibility is to send the negative mass squared of the LSM to infinity
at the integrand level, which leads to a convenient framework for performing OPE computations in the NLSM with the
standard dimensional regularization for the UV divergences. Alternatively, one can also take the limit after
integrating over momenta. In this way, it is the negative mass squared of the LSM that becomes the UV cutoff,
and the situation is then similar to a lattice regularization. We will see that, at least to the next-to-leading order (NLO)
in the $1/N$ expansion, the physics at the cutoff scale decouples
from the NLSM in the IR with a non-perturbative mass scale. We also find that two different perturbative
frameworks -one for the NSLM, another for the LSM at the cutoff scale based on the expansion around the
$O(N)$ broken vacuum~\cite{Jevicki:1977zn,Marino:2019fvu,Marino:2025ido}- naturally connect to
each other in the common region where the momentum becomes much larger than the
mass gap, while much smaller than the cutoff. Moreover, by investigating the possible
appearance of power divergences and their cancellation with ambiguities in the finite parts,
one can also diagnose better the UV/IR nature of the latter. This in particular clarifies further the
physical mechanism behind the cancellation
between the UV renormalon and the condensate of the Lagrangian operator.

The organization of the paper is as follows: in Sec.~\ref{sec-rev} we review the
traditional perturbative framework of the NLSM and its non-perturbative large $N$ solution at
NLO order in the $1/N$ expansion. In particular, we recall that the two-point function, in a convenient
sharp momentum cutoff (SM) scheme, can be expressed in terms of well-defined integrals and expanded as
a trans-series. The goal of  Sec.~\ref{sec:OPE} is precisely to provide a framework, based on a limit of the LSM
at the integrand level, that allows to generate this trans-series by using massless OPE computations
with condensate insertions. A special feature of this framework is that one needs to sum
over infinitely many scale-less condensates in order to obtain the perturbative series at
each order in the 't Hooft coupling. The case of the perturbative self-energy is
discussed in Sec.~\ref{sec:OPEpert}, while the first exponentially small contribution with a scaleful condensate insertion is discussed in Sec.~\ref{sec:OPEcond}. In Sec. \ref{sec:UVrenor} we address in detail the unconventional cancellation
between the UV renormalon and the condensate of the $X$ field, by doing the analysis in a scheme were
power divergences are manifest. In Sec.~\ref{sec:LSM} we discuss what happens when the take the
limit of LSM in a different order, so that the mass of the ``Higgs particle'' becomes an
UV cutoff for the NLSM. In particular, in Sec.~\ref{sec:LSMpert} we show how the perturbative
frameworks for the LSM and NLSM are connected to each other, while in Sec.~\ref{sec:LSMcon} we extend the relationship also to exponentially small contributions. In Sec.~\ref{sec:renorUV} we discuss again the leading renormalon in the perturbative self-energy of the NLSM and its cancellation, this time by using the structure of the power
divergences in the LSM regularization of the NLSM. Finally, we comment and conclude
in Sec.~\ref{sec:concl}. Useful technical results are collected in the two appendices.

\sectiono{The NLSM and its large $N$ solution}
\label{sec-rev}

\subsection{Perturbative approach}
The non-linear sigma model is defined by the bare Euclidean Lagrangian
\be
\CL={1\over 2 g_0 } (\nabla \bss_0 )^2 \ ,
\ee
where the field $\bss_0=(\sigma^1_0(x), \cdots, \sigma^N_0(x))$ satisfies the constraint
\be
\label{constraint}
\bss^2_0(x)=1 \ .
\ee
It will be convenient to introduce a field
\be
\bPhi_0= {1\over g_0^{1/2}} \bss_0 \ ,
\ee
with a canonically normalized kinetic term, and an auxiliary field $X$ to impose
the constraint (\ref{constraint}). In terms of these fields, the Lagrangian reads
\be
\label{lagPhi}
\CL={1\over 2  } (\nabla \bPhi_0 )^2 + {1\over 2} X \left( \bPhi_0^2-{1\over g_0} \right)\ .
\ee
The EOM of this model is
\be
\label{eom}
X= g_0 \bPhi_0 \cdot \nabla^2 \bPhi_0 \ ,
\ee
which will be used later on. The NLSM model has a global $O(N)$ symmetry, acting as rotations of $\bss_0$. It is renormalizable and asymptotically free \cite{polyakov,bzj}.
In this paper we will mostly use the ${\overline{\text{MS}}}$ renormalization scheme and set
\be
g_0 = g Z_g \nu^{\epsilon}, \qquad \bss_0= Z^{1/2} \bss \ ,
\ee
where $\epsilon$ is related to the number of dimensions of the theory by
\be
\label{defeps}
d=2-\epsilon \ ,
\ee
and
\be
\label{nu-def}
\nu^2 = \mu^2 \re^{\gamma_E - \ln(4\pi)} \ .
\ee
 The beta function is defined by
\be
\beta(g; \epsilon)= -{\epsilon g \over 1+ g {\partial \over \partial g} \ln Z_g } \ ,
\ee
and at one-loop one finds \cite{polyakov}
\be
\beta(g;\epsilon)=- \epsilon g-{N-2 \over2 \pi } g^2+ \cdots
\ee
The anomalous dimension of the field $ \bss$ is defined as usual by
\be
\gamma(g)= \beta(g;\epsilon) {\partial \ln Z \over \partial g} \ ,
\ee
and is given at one loop by
\be
\gamma(g)= {N-1 \over 2 \pi} g+ \cdots \ .
\ee

The model can be also analyzed in a $1/N$ expansion, both from a perturbative and a non-perturbative
point of view. The 't Hooft parameter is defined as
 \be
 \label{thooft}
 \lambda_0= {N g_0 \over 2 \pi} \ .
 \ee
Its beta function is given by
\be
\beta_{\lambda} (\lambda;\epsilon)= {N \over 2 \pi }\beta(g;\epsilon)=  -\epsilon \lambda+ \beta_{\lambda} (\lambda) \ .
\ee
Its $1/N$ expansion has the form:
\be
\label{betaN}
\beta_{\lambda} (\lambda;\epsilon) = \sum_{\ell\ge 0} \beta^{(\ell)}_\lambda (\lambda; \epsilon) N^{-\ell} \ .
\ee
We have $\beta^{(\ell)}_\lambda (\lambda; \epsilon) =\beta^{(\ell)}_\lambda (\lambda)$ for $\ell \ge 1$ and
\be
\beta^{(0)}_\lambda (\lambda;\epsilon)=-\epsilon \lambda-\lambda^2 \ .
\ee
The first subleading corrections in $1/N$, but at all loops, can be computed in closed form. One has \cite{bh},
\be
\beta^{(1)}_\lambda(\lambda)=
2 \lambda^2  -4 \lambda^2 \int_0^{\lambda}  \rd x\, \frac{\sin(\frac{\pi x}{2})}{\pi x}\frac{\Gamma(1+x)}{\Gamma(1+\frac{x}{2})^2}  \frac{x+1}{x+2} \ .
\label{beta-bh-1}
\ee
We recall that the renormalization constant of $\lambda$ is given in terms of the beta function by
\be
\label{zlambda}
Z_\lambda= \exp \left[ -\int_0^\lambda {\rd u \over u} {\beta_\lambda(u)\over \beta_\lambda (u; \epsilon)} \right] \ ,
\ee
while the field renormalization constant reads, in terms of the anomalous dimension $\gamma(\lambda)$, as
  \begin{align}\label{zfield}
  Z=\exp \left[\int_0^{\lambda}\frac{  \gamma(u)}{ \beta_{\lambda}(u; \epsilon)}{\rd u } \right] \ .
  \end{align}
The anomalous dimension also has a $1/N$ expansion of the form
\be
\gamma(\lambda) = \sum_{\ell \ge 0} \gamma^{(\ell)} (\lambda) N^{-\ell} \ .
\ee
The leading term is given simply by
\be
\gamma^{(0)}(\lambda)= \lambda \ ,
\ee
while the subleading term is (see e.g. \cite{cr-dr})
\be
\gamma^{(1)}(\lambda)=   {\lambda (2-\lambda) \sin \left(\frac{\pi  \lambda}{2}\right) \Gamma (\lambda) \over \pi  \Gamma \left(1+\frac{\lambda}{2}\right) \Gamma
   \left(2+\frac{\lambda}{2}\right)} - {\beta^{(1)} (\lambda) \over \lambda} \ .
   \ee
   We will rederive this result as a spinoff of our method and calculations.

Let us consider now the anomalous dimension of the Lagrangian operator of the theory, which has
been studied in \cite{montanari-thesis, montanari-paper}. This operator mixes with the operator\footnote{In \cite{blgzj,montanari-thesis, montanari-paper} IR divergences are regularized by introducing an external field $H$, and this leads to an $H$-dependent correction to the operator $\alpha$. Since we only consider $O(N)$ invariant quantities in this work, all the IR divergences cancel in the end~\cite{ElitzurIR,Daniel-Gabriel,DavidIR}, so we can set $H$ to zero.}
\be
\alpha= {\nabla^2 \sigma^N_0 \over \sigma^N_0}
\ee
and we have the following matrix equation for their renormalization:
 \be
 \begin{pmatrix} ~(\partial \bss_0)^2 \\ \alpha \end{pmatrix}= \mZ
 \begin{pmatrix} ~[(\partial \bss)^2] \\ [\alpha] \end{pmatrix},
 \ee
where $[\cdot]$ denotes renormalized operators, and
\be
\mZ=Z_g \begin{pmatrix}  1- {\beta(g) \over g \epsilon} - {\gamma (g)  \over \epsilon} & -\frac{\gamma (g)}{\epsilon } \\[2ex]
 {\beta(g) \over g \epsilon} + {\gamma (g)  \over \epsilon} & 1+{\gamma (g) \over \epsilon }
 \end{pmatrix}.
 \ee
By using the EOM one can show that $\alpha= -(\partial \bss_0)^2$. Therefore, when considering $O(N)$-invariant
vacuum expectation values, we have the simpler renormalization,
\be
\label{ren-op}
(\partial \bss_0)^2= Z_\lambda \left(1- { \beta_\lambda (\lambda) \over \lambda \epsilon} \right) [(\partial \bss)^2] \ ,
\ee
which we have already expressed in terms of the 't Hooft coupling.
By using (\ref{zlambda}) and manipulations similar to the ones in \cite{tsc} one finds the $1/N$ expansion
\be
Z_\lambda \left(1- { \beta_\lambda (\lambda) \over \lambda \epsilon} \right)=1+ {1\over N} \int_0^\lambda {\rho(u) \over u(u+\epsilon)} \rd u+ \CO\left( N^{-2})\right) \ ,
\ee
where
\be
\rho(\lambda)= -\lambda^2  {\rd \over \rd \lambda}\left( {\beta^{(1)}_\lambda(\lambda) \over \lambda^2} \right)= 4 \lambda \frac{\sin(\frac{\pi \lambda}{2})}{\pi}\frac{\Gamma(1+ \lambda)}{\Gamma(1+\frac{ \lambda}{2})^2}  \frac{ \lambda+1}{ \lambda+2} \ .
\ee
The anomalous dimension of the operator $(\partial \bss)^2$ can also be obtained by noticing that the operator $\frac{\beta}{2g^2}[(\partial \bss)]^2$ is proportional to the trace of the energy-momentum tensor $T^{\mu\nu}$ and does not renormalize. Therefore, it can be extracted as
\be
\frac{\beta(g)}{2g^2}\frac{\rd}{\rd\ln \mu}
\left(\frac{2g^2}{\beta(g)}\right)=\frac{2}{g}\beta(g)-\beta'(g)=-g^2\frac{\rd}{\rd g} \left(\frac{\beta(g)}{g^2}\right) \ ,
\ee
which agrees with the derivations above.

As a spinoff of our calculation we will derive the value of the renormalization constant (\ref{ren-op}),
which leads in turn to a simple diagrammatic method to obtain the value of the beta function at NLO in $1/N$.
This is similar to the calculation in the Gross--Neveu case in \cite{tsc}. As a general rule, OPE/condensate calculations turn out to be useful to determine anomalous dimensions, as emphasized recently in \cite{opead}.

\subsection{Exact large $N$ approach}

As is well-known (see e.g. \cite{mmbook} for a review), the NLSM can be solved at large $N$, starting with the
representation (\ref{lagPhi}). There is a non-trivial large $N$ saddle-point where the
auxiliary field $X$ becomes dynamical and acquires a vev. At leading order in $1/N$ one finds
\be
\langle X \rangle_0 = m_0^2 \ ,
\ee
where $m_0$ is determined by the gap equation
\be
\label{gapeq}
{1\over N g_0}= \int {\rd^d k \over (2 \pi)^d} {1\over k^2+ m_0^2} \ .
\ee
The vev of $X$ gives a mass to the $\bPhi$ field. The propagator for the fluctuations of the $X$ field is given by
 \be
 \label{propa-X}
 \langle X(k)X(-k) \rangle=\frac{1}{N}\Delta (k; m_0)= {4 \pi k^2 \xi  \over N} \left[\ln \left( {\xi+1 \over \xi-1} \right) \right]^{-1} \ ,
 \ee
where
\be \label{eq:defxi}
 \xi = {\sqrt{1+ {4 m_0^2 \over k^2} }} \ .
 \ee
Therefore, the large $N$ theory describes an $X$ particle interacting with the $N$ scalar fields $\bPhi$, in which the coupling is of order $N^{-1/2}$.
The correlation functions in this theory can be computed in a systematic expansion in $1/N$, by means of large $N$ Feynman diagrams. In contrast
to conventional perturbative series, each coefficient
in the $1/N$ expansion is, after renormalization, an exact function of the 't Hooft coupling.

\begin{figure}
\centering
\includegraphics{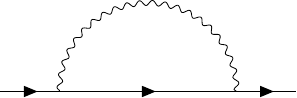}
\qquad \qquad
\includegraphics{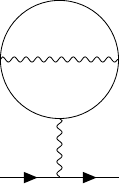}
\caption{Diagrams contributing to the two-point function
of the NLSM at large $N$. We will refer to the diagram on the left as the self-energy diagram, and the one on the
right as the tadpole diagram. The wavy line corresponds to the large $N$ propagator of the $X$ field. }
\label{setad-fig}
\end{figure}%

The self-energy of the $\bPhi$ field, $\Sigma(p)$, is defined by
\be
\label{se-def}
{1\over N} \langle \bPhi(p) \cdot \bPhi(-p)\rangle= {1\over p^2+ m_0^2 + {1 \over N} \Sigma(p, m_0^2)} \ ,
\ee
and it has the $1/N$ expansion
\be
\Sigma(p, m_0^2)=\Sigma^1(p, m_0^2)+ \CO(1/N) \ ,
\ee
where $\Sigma^1(p, m_0^2)$ can be calculated by standard methods in the exact large $N$ expansion.
It is given by (see e.g. \cite{cr-dr})
\be
\label{sigma1}
\Sigma^1(p, m_0^2)=\Sigma^1_{S} (p, m_0^2)+\Sigma^1_{T}(m_0^2) \ ,
\ee
where
\be
\label{NLSM-ST}
\ba
\Sigma^1_{S} (p, m_0^2)&=\int {\rd^d k \over (2 \pi)^d} {\Delta (k;m_0) \over (p+k)^2+ m_0^2} \ , \\
\Sigma^1_{T}(m_0^2)&= {\Delta(0;m_0) \over 2}  \int {\rd^d k \over (2 \pi)^d}\Delta(k;m_0) {\partial \over \partial m_0^2} \Delta^{-1}(k;m_0) \   .
\ea
\ee
are the contributions of the self-energy diagram and the tadpole diagram, respectively. These diagrams are shown in \figref{setad-fig}.
These expressions are UV divergent and they need regularization and renormalization. We will follow \cite{cr-dr}, who use a sharp
momentum cutoff (SM) regularization scheme. In this scheme, one first performs the integration over angular variables
in (\ref{sigma1}). Then, the integrand which is obtained in this way is Taylor expanded at infinity. The terms which lead to a
divergence are simply subtracted, but in order to avoid IR divergences in the
subtracted pieces one has to introduce an IR cutoff $M$, which plays the role of the renormalization scale $\mu$ in dimensional regularization, see \cite{cr-review,biscari} for more details.

After this procedure is implemented, one finds a finite two-point function calculated in \cite{cr-dr}. We will now present a slightly different derivation of this function, which will be useful for our purposes. Let us first consider the contribution of the tadpole diagram. We first notice that
the integrand for $\Sigma^1_T$ can be simplified as
\begin{align}
\frac{\Delta(0,m_0^2)}{2}\Delta(k;m_0) {\partial \over \partial m_0^2} \Delta^{-1}(k;m_0)=-\frac{\Delta(k^2,m_0^2)}{k^2+4m_0^2}-\frac{8\pi m_0^2}{k^2+4m_0^2} \ .
\end{align}
Thus, by using SM regularization we find
\be
\label{sigma-I}
\Sigma^1_T(m_0^2, M^2)=-2m_0^2\ln \left( \frac{M^2}{4m_0^2} \right) +{\cal I}\, ,
\ee
where
\be
{\cal I}=\left[-\int_0^{\Lambda^2} \frac{ \rd k^2}{4\pi } \frac{\Delta(k^2;m_0^2)}{(k^2+4m_0^2)}+\int_{M^2}^{\Lambda^2} \rd k^2 \left( \frac{1}{\ln \left( \frac{k^2}{m_0^2} \right) }- {2 m_0^2 \over k^2} \frac{1+\ln  \left( \frac{k^2}{m_0^2} \right)}{\ln^2 \left( \frac {k^2}{m_0^2}\right)}\right)\right]_{\Lambda^2 \rightarrow \infty}\ .
\ee
In this expression we have subtracted, in the second term, the large $k^2$ expansion of the integrand in the first term, up to the order of convergence. Then, by introducing the Borel parameter for the inverse logarithm when $A>1$
\begin{align}
\frac{1}{\ln A}=\int_0^{\infty} \rd t A^{-t} \ ,
\end{align}
one ends up with the following representation for ${\cal I}$
\begin{align}
{\cal I}=-\lim_{\Lambda^2\rightarrow \infty}\int_0^{\infty}
\rd t \bigg(\int_0^{\Lambda^2} \frac{\rd k^2}{\xi} \left(\frac{\xi+1}{\xi-1}\right)^{-t}-\int_{M^2}^{\Lambda^2} \rd k^2 \left(\frac{k^2}{m_0^2}\right)^{-t}\left(1-\frac{2m_0^2(1+t)}{k^2}\right) \bigg) \ ,
\end{align}
where $\xi$ is defined in Eq.~(\ref{eq:defxi}). Since the large $k^2$ singularities are subtracted uniformly for $0<t<1$ (for $t>1$ the integrals over $k^2$ are automatically convergent), the limit $\Lambda^2 \rightarrow \infty$ can be taken inside the  integral over $t$. It defines an analytic function in $t$ which can be obtained by analytic continuation from the region ${\rm Re}(t)>1$
\begin{align}
&-\lim_{\Lambda^2 \rightarrow \infty}\bigg(\int_0^{\Lambda^2} \frac{\rd k^2}{\xi} \left(\frac{\xi+1}{\xi-1}\right)^{-t}-\int_{M^2}^{\Lambda^2} \rd k^2 \left(\frac{k^2}{m_0^2}\right)^{-t}\left(1-\frac{2m_0^2(1+t)}{k^2}\right) \bigg) \nonumber \\
&=\frac{M^2}{t-1}\left(\frac{M^2}{m_0^2}\right)^{-t}-\frac{2m_0^2(1+t)}{t}\left(\frac{M^2}{m_0^2}\right)^{-t}+\frac{2m_0^2}{t(1+t)(1-t)} \ .
\end{align}
This function has no singularities when $t>0$ and is integrable at large $t$ if $M^2>m_0^2$. Thus, by combining it with the first term in the r.h.s. of (\ref{sigma-I}), and after changing variables $t=y/2$, one obtains
\be
\label{smt}
\Sigma^1_T(m_0^2, M^2)=M^2\int_0^{\infty}  \frac{1}{y-2}\left(\frac{M}{m_0}\right)^{-y}\rd y+\widehat m_1^2,
\ee
where
\be
\ba
\widehat m_1^2&=-2m_0^2\ln \left( \frac{M^2}{4m_0^2}\right) +2m_0^2\ln \ln \left( \frac{M^2}{m_0^2} \right)-\frac{2m_0^2}{\ln \left( \frac{M^2}{m_0^2}\right)}\\
&+2m_0^2\int_0^{\infty}\bigg(\frac{4}{y(2+y)(2-y)}  -\frac{\re^{-y/2}}{y}\bigg)\rd y \ .
\ea
\ee
The result (\ref{smt}) has a power divergence in $M^2$, multiplying an integral with a pole at $y=2$. This pole is cancelled by the one of $\widehat m_1^2$, so the total expression is finite.
By using the $1/N$ correction to the free energy $F(m)$ calculated in \cite{biscari, cr-dr}, it can be checked that the finite part of $\widehat m_1^2$ is the $1/N$ correction to the vev of $X$.

Let us now consider the self-energy diagram. After SM regularization, we obtain
\be
\ba
\Sigma_{S}^1 (p^2, m_0^2, M^2)&=\Sigma^1_{\rm SM}(p, m_0^2)+  \int_0^\infty  \left( \ln {\xi +1 \over \xi-1} \right)^{-1} \left( 1- {p^2+m_0^2 \over 2 m_0^2} \left({1\over \xi} -1\right)\right)\rd k^2\\
& -  \int_{M^2}^\infty \left( {1\over \ln(k^2/m_0^2)} + {p^2 +m_0^2 \over k^2 \ln(k^2/m_0^2)}-{2 m_0^2  \over k^2  \ln^2(k^2/m_0^2)} \right)\rd k^2.
\ea
\ee
In this equation, $\Sigma^1_{\rm SM}(p, m_0^2)$ is the renormalized $1/N$ correction to the self-energy:
 \be
 \label{sigmaSM}
\Sigma_{\rm SM}^1(p, m_0^2)= m_0^2 S\left( {p^2 \over m_0^2} \right) \ ,
 \ee
 where
 \be
 \label{Sx}
 S(x)= \int_0^\infty \rd y \left[ \ln\left( {\xi+ 1  \over \xi-1}  \right)\right]^{-1} \left[  {y  \xi \over {\sqrt{(1+ y +x)^2-4 x y}}} -1+
  {x+ 1 \over 2} \left( {1\over \xi}-1\right) \right] \ .
  \ee
  This function has the property that $S(-1)=0$ \cite{cr-dr}. Proceeding as before, we obtain the representation
\be
\label{ssreg-t}
\ba
\Sigma_{S}^1 (p^2, m_0^2, M^2)&= -M^2 \int_0^{\infty}  \frac{1}{y-2}\left(\frac{M}{m_0}\right)^{-y}\rd y+4 m_0^2 \int_0^\infty {\rd y \over (y+2)(y-2)} + \frac{2m_0^2}{\ln \left( \frac{M^2}{m_0^2}\right)} \\
& + (p^2 + m_0^2)  \left(  \ln \ln \left( {M^2 \over m_0^2}\right)+\gamma_E  \right) + \Sigma_{\rm SM}^1(p, m_0^2).
\ea
\ee
The first term in the r.h.s. of (\ref{ssreg-t}) has a power divergence equal and opposite to the one in (\ref{smt}).
The pole at $y=2$ appearing in this term cancels against a similar pole in the second term.
By adding (\ref{smt}) and (\ref{ssreg-t}) one
finds the result in \cite{cr-dr},
\be
\label{sigma1-ren}
 \Sigma^1(p)=\Sigma^1_{\rm SM}(p, m_0^2)+ (p^2+ m_0^2) \left( \ln\ln\left( {M^2 \over m_0^2} \right)+ \gamma_E \right)+ m_1^2 \ .
 \ee
 In this expression,
\be \label{eq:SMmass1}
m_1^2=2 m_0^2
 \left(-\ln\left(  {M^2 \over 4 m_0^2} \right)+\ln\ln\left( {M^2 \over m_0^2} \right)+ \gamma_E \right) \ ,
 \ee
 is the $1/N$ correction to the physical mass gap
 \be
 m^2= m_0^2+ {1\over N} m_1^2+ \CO( N^{-2}) \ .
 \ee
 We note that the power divergences cancel in the final expression, a consequence of the multiplicative renormalizability of the theory. We also note that, in the SM scheme, one has \cite{cr-dr}
\be
\beta_\lambda^{\rm SM} (\lambda)= - \lambda^2 + {1\over N} \left(2\lambda^2- \lambda^3 \right)+ \CO\left(N^{-2} \right),
\qquad \gamma^{\rm SM}(\lambda)= \lambda+ {1\over N} \left( -\lambda+ \lambda^2 \right)
+ \CO\left(N^{-2} \right) \ .
\ee

The asymptotic expansion of $\Sigma^1_{\rm SM}(p, m_0^2)$ at large $p^2$ should agree with the perturbative series for
the self-energy obtained in massless perturbation theory. This is implicit in the results of \cite{cr-dr, beneke-braun} and will be
verified explicitly in the next section. It was found in \cite{beneke-braun} that
$\Sigma^1_{\rm SM}(p,m_0^2)$ has in fact a trans-series expansion
including exponentially small corrections. Let us review this result. We first note that, at the order we are working in
$1/N$, we can replace $m_0^2$ by $m^2$, and the running 't Hooft coupling at the scale set by $p$ is given by
\be
{ m^2 \over p^2}= \re^{-2/\lambda(p)} \ .
\ee
Then, one has the representation
\begin{equation}
\label{fin-int}
{m^2 \over p^2} S\left({p^2 \over m^2} \right)=\sum_{n \ge 0} (-1)^n \re^{-2n/\lambda} \int_0^\infty \left\{  \re^{-y/\lambda} \left( \frac{F_n(y)}{\lambda} + G_n(y) \right) -H_n(y) \right\}\rd y \ .
\end{equation}
Explicit expressions for the functions $F_n$, $G_n$ and $H_n$ were obtained in
\cite{beneke-braun} in a slightly different renormalization scheme. The results in the SM scheme considered here are very similar and
can be found in \cite{mmr-ts}, after changing the variable $t$ used in
\cite{beneke-braun, mmr-ts} to $y/2$. In this paper we will only need the explicit result for
$n=0$,
\be
\label{n0fgh}
F_0(y)=1 \ , \qquad G_0(y)= - \gamma_E - {1\over 2} \psi\left({y \over 2}-1\right) -{1\over 2}
\psi\left(2-{y \over2} \right), \qquad H_0(y) =  \frac{1}{y} - \frac{1}{y+2}\ ,
\ee
and for $n=1$,
\be
\label{n1fgh}
\ba
F_1(y)&={y^2 \over 4}-1, \\
G_1(y)&=-{1\over y} +{1\over 2} -y+ {y^2 \over4 } + \frac{1}{2} \left(1-\frac{y^2}{4}\right) \left(\psi\left(2-\frac{y}{2}\right)+\psi
  \left(\frac{y}{2}+1\right)+2 \gamma_E \right) , \\
H_1(y) &=  -\frac{1}{y} - \frac{1}{y-2} + \frac{2}{y+2} \ .
\ea
\ee
These functions have singularities on the positive real axis. Although the singularities
cancel among different terms in the sum over $n$ in (\ref{fin-int}), it is convenient to write this
expression as a Borel-resummed trans-series. The first step is to remove the singularities at the origin,
which are cancelled between the functions $G_n(y)$ and $H_n(y)$. Let us define
\be
r_{n,k} = {\rm Res}_{y=2k} \, H_n(y) \ ,
\ee
and introduce the function
\be
\widehat G_n(y)= G_n(y) - \frac{r_{n,0}}{y} \ ,
\ee
which is regular at $y=0$. Next, we deform the integration contour in (\ref{fin-int}) slightly above or below the positive real axis, so that the integral is well-defined. Then, we can express it in term of lateral Borel resummations of factorially divergent series,
 \be
 \int_0^{\infty \re^{\ri\theta_\pm}} \re^{-y/\lambda} \left( \frac{F_n(y)}{\lambda} + \widehat G_n(y) \right)  \rd y = s_\pm \left( \varphi_n \right)(\lambda) \ ,
 \ee
 where $\theta_\pm$ is a small positive (respectively, negative) angle and
\be
\label{n-trans}
\varphi_n(\lambda)= F_n(0)+ \sum_{k \ge 0} \left( F_n^{(k+1)}(0) + \widehat G_n^{(k)}(0)  \right) \lambda^{k+1} \ .
 \ee
Let us now consider the last term in the integral (\ref{fin-int}). A straightforward calculation gives,
\be
\label{hn}
-\int_0^{\infty \re^{\ri\theta}} \left( H_n(y) - r_{n,0} \frac{\re^{-y/\lambda}}{y} \right)\rd y =
r_{n,0}\ln(\lambda) + c_n  \pm \ri\pi \sum_{k=1}^n r_{n,k} \ ,
\ee
where the constant $c_n$ is defined as the principal value integral
\begin{equation}
\label{cn}
c_n = -{\rm{P}} \int_0^\infty \left( H_n(t) -r_{n,0} \frac{ \re^{-t}}{t} \right) \rd t \ ,
\end{equation}
and in the last term of the r.h.s. of (\ref{hn}) we pick the residues of $H_n(y)$ along the positive real axis,
with a sign depending on the deformation of the contour.

It is convenient to consider trans-series including not only the power series (\ref{n-trans}), but also the constant and logarithmic terms arising in (\ref{hn}), (\ref{cn}). We then define,
\be
\label{nt-trans}
(-1)^n \Phi_n(\lambda)= r_{n,0}\ln(\lambda) + c_n  \pm \ri\pi \sum_{k=1}^n r_{n,k}+ \varphi_n(\lambda) \ .
\ee
When $n=0$, we have
\begin{equation}
\label{Phi0}
\Phi_0(\lambda)= \ln\left(\lambda/2\right)- \gamma_E +1-\lambda+ \sum_{k \ge 1 }a_{0,k} \lambda^{k+1} \ ,
\ee
where
\be
a_{0,k}=2^{-k-1} k! \left( \left(1+(-1)^k\right)\zeta(k+1) - 2 \right), \qquad k \ge 1.
\end{equation}
The result for $n=1$ is
\be
\label{Phi1}
\Phi_1(\lambda)=\ln\left(\lambda/2\right)- \gamma_E \pm \pi \ri +1- \lambda+ {\lambda^2 \over 4}  + { \zeta(3)-1\over 2} \lambda^3+\sum_{k \ge 2} a_{1,k} \lambda^{2k+1},
\ee
where
\be
a_{1,k}= 2^{-2k} (2k)! (\zeta(2k+1)-\zeta(2k-1)) \ .
\ee
%

%
%
%

\section{The NLSM as a limit of the LSM} \label{sec:OPE}

\subsection{Perturbative self-energy}\label{sec:OPEpert}
As we mentioned in the Introduction, the constraint (\ref{constraint}) leads to technical difficulties in the treatment
of the NLSM. One possibility is to solve the constraint and express the $N$-th component of the field in terms of
the first $N-1$ components,
as done in e.g. \cite{blgzj}. This leads to a model with an infinite number of interaction vertices, and even proving
its renormalizability requires some ingenuity. To overcome these difficulties, and in particular
to reproduce the results of \cite{beneke-braun} from a condensate calculation,
we will appeal to the old idea \cite{bessiszj,bzj} that the NLSM can be obtained as a certain limit of the {\it linear} sigma model
with a quartic interaction.

Consider then the following action for the fields $\bPhi_0$ and $D$, given by
\begin{align}
S=\int \rd^dx \bigg(\frac{1}{2}(\nabla \bPhi_0)^2-\frac{\ri D }{2}\left(\bPhi_0^2-\frac{N}{2\pi \lambda_0}\right)+\frac{D^2}{2\Lambda^2}\bigg) \ . \label{eq:deflinear}
\end{align}
If we integrate out $D$ we obtain the quartic LSM with a negative squared mass.
We can also regard $\Lambda \gg \mu$ as an artificial UV scale for the $D$ field. The NLSM is then obtained as
the limit $\Lambda \rightarrow \infty$ of the theory defined by (\ref{eq:deflinear}), {\it before} $\epsilon \rightarrow 0$, and at the integrand level. This means that, when integrating over momenta, one has to implement this limit in the integrand before performing the integration. By taking this limit we suppress all the
contributions at the scale $\Lambda^2$, and only the asymptotically free, short distance limit of the NLSM is left\footnote{If one reverses the order of limits, as we will discuss in section \ref{sec:LSM}, then $\Lambda$ becomes a semi-hard scale that plays the role of an UV cutoff for the NLSM.}. The Feynman rules of the massless NLSM are then
\begin{align}
\langle D(p)D(-p)\rangle=\Lambda^2  \ , \\
\langle \bPhi^a_0(p)\bPhi^{b}_0(-p)\rangle=\frac{\delta^{ab}}{p^2} \ ,
\end{align}
and the triple interaction vertex is $\ri/2$. Let us note that, when taking the limit to the NLSM, the field
$D$ becomes $\ri X$ in (\ref{lagPhi}). In Feynman diagrams, we will represent the $D$ field by a dashed black line, while the $\bPhi_0$ field will be represented by a continuous line. In the computation of coefficient functions, sub-diagrams that are forced to be scaleless (such as tadpoles of the $D$-field) are set to zero, according to the standard rules of dimension regularization. Intuitively, these sub-diagrams receive only low-momentum contributions and are added back through operator condensates in the OPE (see e.g. \cite{pascual-tarrach} for a discussion of this point). 

One of our goals in this paper is to show in detail how the two-point function
of the NLSM can be obtained by using an OPE computation in the limit
$\Lambda^2 \rightarrow \infty$ of the linear sigma model (\ref{eq:deflinear}).
In this section we will first consider the perturbative two-point function of the NLSM,
and in the next section we will consider the non-perturbative corrections. In both
cases we will be able to match the results obtained in the exact large $N$ calculation
reviewed in section \ref{sec-rev}, after taking into account the change of scheme.

An important feature of our computation is the manifest $O(N)$ invariance. We will assume that the ground state supports only $O(N)$ invariant condensates, such as
\begin{align}
&\langle \bPhi^{a_1} \bPhi^{a_2}\rangle =\frac{\delta^{a_1a_2}}{N}\bPhi \cdot \bPhi \ , \\
&\langle \bPhi^{a_1} \bPhi^{a_2} \bPhi^{a_3}\bPhi^{a_{4}} \rangle=\frac{1}{N^2}\left(\delta^{a_1a_2}\delta^{a_3a_4}+\delta^{a_1a_3}\delta^{a_2a_4}+\delta^{a_1a_4}\delta^{a_2a_3}\right)\langle \left( \bPhi \cdot \bPhi\right)^2 \rangle \ .
\end{align}
Since the Feynman rules are also $O(N)$ invariant,  all the correlators in our formalism are automatically decomposed in terms of $O(N)$ invariant combinations. In this paper we will focus on the self-energy defined in (\ref{se-def}), which is manifestly $O(N)$ invariant. In the traditional framework of NLSM, such as~\cite{blgzj,montanari-thesis, montanari-paper}, it is common to use an external field $H$ to regulate IR divergences, and it is known that for $O(N)$ invariant quantities, the $H \rightarrow 0$ limit is free from IR divergences~\cite{Jevicki:1977zn,ElitzurIR,DavidIR}. However, as pointed in e.g.~\cite{Daniel-Gabriel}, if one considers only $O(N)$ invariant quantities, it is simpler to use dimensional regularization both for UV and IR divergences, and IR divergences will cancel in the end. As such, we will adopt this approach in our work.

\begin{figure}[!ht]
\leavevmode
\begin{center}
\includegraphics[height=3cm]{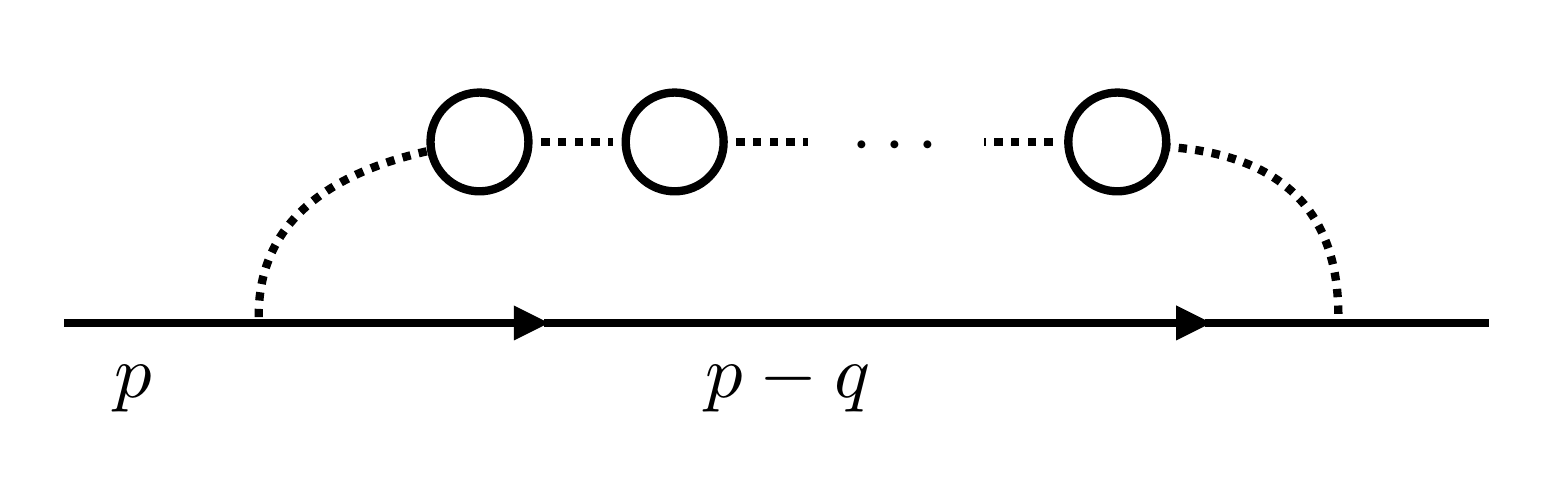}
\end{center}
\caption{Diagram with $n$ insertions of the $\bPhi_0$ bubble in the linear sigma model.}
\label{bubble-fig}
\end{figure}

 Let us consider then the perturbative two-point function of the NLSM and the corresponding self-energy.
The contribution of a bubble of the $\bPhi_0$ field in a Feynman diagram gives
\be
\Pi(p^2, \epsilon)=\frac{1}{2}\int \frac{\rd^d q}{(2\pi)^d}\frac{1}{(p-q)^2 q^2}=-{1\over 2 \pi \epsilon p^2} \left({p^2 \over 4 \pi} \right)^{-\epsilon/2}f(\epsilon) \ ,
\ee
where
\be\label{eq:fe}
f(\epsilon)={ \Gamma \left(1-{\epsilon \over 2}\right) \Gamma \left(1+{\epsilon \over 2} \right)  \Gamma \left(-{\epsilon \over 2} \right)  \over 2 \Gamma \left(-\epsilon \right)} \ .
\ee
The first contribution to the self-energy comes from the diagram considered in
\figref{bubble-fig}, in which we insert $n$ bubbles of the
$\bPhi_0$ field. This can be regarded as the coefficient function for the identity
operator for the self-energy. In total, we have
\be
C_{\rm I}(p^2; \Lambda^2)=\sum_{n=0}^{\infty}
(-1)^n (\Lambda^2)^{n+1}N^n \int\frac{\rd^d q}{(2\pi)^d} {\Pi^n(q^2,\epsilon)\over (p-q)^2} =\int\frac{\rd^d q}{(2\pi)^d}\frac{1}{(p-q)^2}\frac{\Lambda^2}{1+\Lambda^2 N \Pi(q^2,\epsilon)} \ .
\ee
According to the rules stated above, we have to take the limit
$\Lambda^2 \rightarrow \infty$ at the integrand level, which leads to
\be
C_{\rm I}(p^2)={1\over N} \int\frac{\rd^d q}{(2\pi)^d}\frac{1}{(p-q)^2}\frac{1}{\Pi(q^2,\epsilon)} \ . \label{eq:CINLSM}
\ee
In this way the unphysical $k^2\approx\Lambda^2$ region drops, and one is left with the
desired massless NLSM integrals. However, in order to obtain the perturbative propagator of the NLSM
one has to impose the constraint (\ref{constraint}), which in terms of $\bPhi_0$ reads
\be
{1\over N} \bPhi_0^2= {1\over 2 \pi \lambda_0}. \label{eq:constraint1}
\ee
We can regard this as a condensate for the operator $\bPhi_0^2$, and we can incorporate
the constraint by considering diagrams with insertions of condensates of $\bPhi_0^2$. A similar strategy
has been used to incorporate vevs of scalar fields in \cite{scalar-ope}. Equivalently,
we have to include the following bare operators in the OPE
\begin{align}
{\cal O}_{0,\ell}=\left(\frac{1}{N}\bPhi_0^2\right)^\ell \ .
\end{align}
\begin{figure}[!ht]
\leavevmode
\begin{center}
\includegraphics[height=6cm]{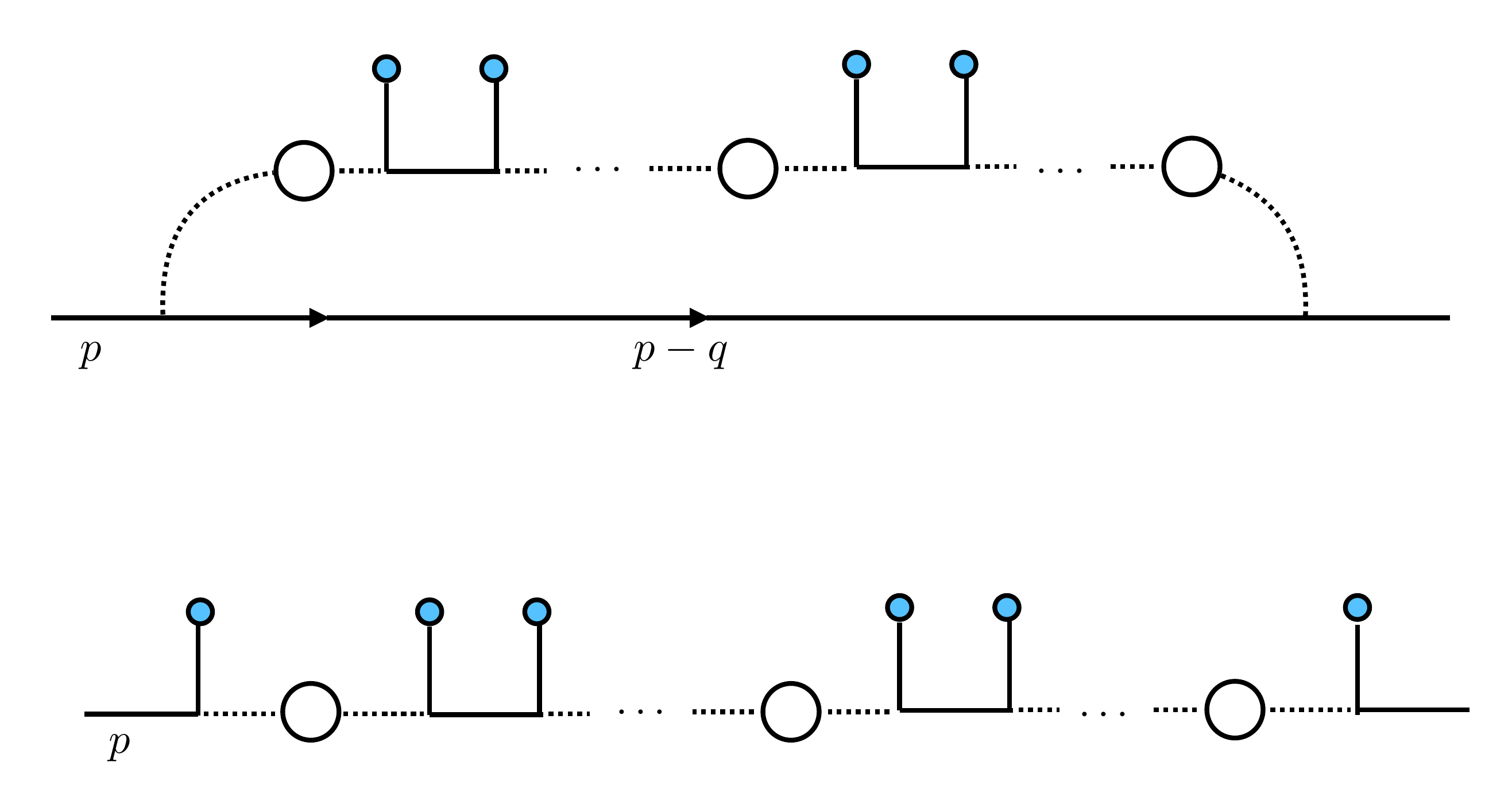}
\end{center}
\caption{Diagrams with $n-\ell$ insertions of $\bPhi_0$ bubble and $\ell$ insertions of the $\bPhi_0^2$ condensates.}
\label{more-bubble-fig}
\end{figure}
At this order in $1/N$, only the tree-level condensates for them are required\footnote{In this work, we only use the bare version of these scaleless operators, as their main role is to introduce, through the constraint Eq.~(\ref{eq:constraint1}), the bare coupling-constant $\lambda_0$  that will be renormalized latter in a standard manner. In particular, operator mixings among them are encoded by the coupling constant renormalization and need no special treatment in our approach.}. The condensates of the
$\bPhi_0$ field will be represented in Feynman diagrams by a
blue blob. Their
coefficient functions can be computed from two sets of diagrams, which are shown
in \figref{more-bubble-fig}. They are similar in structure to the tree-level condensate insertions in the short distance limit
of the quartic model~\cite{Liu:2025edu}. In the first set, we consider the self-energy
one-loop diagram with $n-\ell$ insertions of $\bPhi_0$ bubbles, and $\ell$
insertions of condensates. Its contribution is given by
\begin{align}
C_{0,\ell}^{(a)}(p^2; \Lambda^2)&=\sum_{n=\ell}^{\infty}(-1)^n(\Lambda^2)^{n+1}\binom{n}{\ell}
\int\frac{\rd^d q}{(2\pi)^d}\frac{1}{(p-q)^2}\left(\frac{1}{q^2}\right)^\ell
\left(N \Pi(q^2,\epsilon)\right)^{n-\ell}\nonumber \\
&=\int\frac{d^d q}{(2\pi)^d}\frac{1}{(p-q)^2}\left(\frac{1}{ q^2}\right)^\ell \frac{(-1)^\ell (\Lambda^2)^{\ell+1}}{(1+N\Lambda^2 \Pi)^{\ell+1}} \ .
\end{align}
In the limit $\Lambda^2 \rightarrow \infty$ we find
\be
C_{0,\ell}^{(a)}(p^2)=
\int\frac{\rd^d q}{(2\pi)^d}\frac{1}{(p-q)^2}\left(\frac{1}{q^2}\right)^\ell \frac{(-1)^\ell }{( N \Pi(q^2,\epsilon))^{\ell+1}} \ . \label{eq:O0KNLSP}
\ee
Notice that these coefficient functions are expressed in terms of massless integrals. Now, one
uses the large $N$ factorization
\begin{align}
\langle{\cal O}_{0,\ell} \rangle=\left(\frac{1}{2\pi \lambda_0}\right)^\ell \ ,
\end{align}
to write
\be
\label{bub+cond}
\ba
C_{\rm I}(p^2)+\sum_{\ell=1}^{\infty}C^{(a)}_{0,\ell}(p^2)N^{\ell} \langle{\cal O}_{0,\ell} \rangle&={1\over N} \sum_{\ell=0}^{\infty}\int\frac{\rd^d q}{(2\pi)^d}\frac{1}{(p-q)^2}\left(\frac{1}{q^2}\right)^\ell\frac{(-1)^\ell}{( \Pi (q^2,\epsilon))^{\ell+1}} \left(\frac{1}{2\pi \lambda_0}\right)^\ell \\
&={1 \over N} \int\frac{\rd^d q}{(2\pi)^d}\frac{q^2 }{(p-q)^2}\frac{2 \pi \lambda_0}{1+ 2 \pi \lambda_0 q^2 \Pi(q^2,\epsilon)} \ . \ea
\ee
%
The second set of diagrams is shown in the lower part of \figref{more-bubble-fig}. They have again $n-\ell$ bubbles of the $\bPhi_0$ field, and
$\ell$ condensate insertions. Their contribution is
\be
C^{(b)}_{0,\ell+1}(p^2; \Lambda^2) =\sum_{n=\ell }^{\infty}(-1)^n(\Lambda^2)^{n+1}\binom{n}{\ell}\left(\frac{1}{p^2}\right)^\ell \left(N  \Pi (p^2,\epsilon)  \right)^{n-\ell} \ ,
\ee
so that when $\Lambda^2 \rightarrow \infty$ we find
\be
C^{(b)}_{0,\ell+1}(p^2)=\left(\frac{1}{p^2}\right)^\ell \frac{(-1)^\ell }{( N \Pi (p^2,\epsilon))^{\ell+1}} \ . \label{eq:NLSMCb0k}
\ee
Thus, one has
\begin{align}
\label{chain-sum}
\sum_{\ell=0}^{\infty}C^{(b)}_{0,\ell+1}(p^2)N^{\ell} \langle{\cal O}_{0,\ell+1}\rangle ={p^2 \over 1+ 2 \pi \lambda_0 p^2 \Pi(q^2,\epsilon)} \ .
\end{align}
The total contribution can be written as a an infinite power series in the bare 't Hooft coupling,

\be
\ba
\label{sigma-pert-n}
\Sigma_1^{\overline{\rm MS}}(p^2,\epsilon)&=\sum_{n=0}^{\infty}(2 \pi \lambda_0)^{n+1} \int\frac{\rd^d q}{(2\pi)^d}\frac{ (q^2)^{n+1} \left(-\Pi (q^2, \epsilon) \right)^n }{(p-q)^2} +\sum_{n=0}^{\infty} (2 \pi \lambda_0)^n (p^2)^{n+1} \left(-\Pi (p^2, \epsilon) \right)^n\\
& \equiv p^2 + p^2 \sum_{n=1}^{\infty}  \Sigma_{1,n}^{\overline{\rm MS}}(p^2,\epsilon)\lambda_0^n \ ,
\ea
\ee
and one can use standard one-loop integrals to obtain
\be
 \Sigma_{1,n}^{\overline{\rm MS}}(p^2,\epsilon)= \left( {p^2 \over 4 \pi} \right)^{-n\epsilon/2} \left( {f(\epsilon) \over \epsilon} \right)^{n-1} \left\{ {f(\epsilon) \over \epsilon} +  \frac{\Gamma \left(2-{n\epsilon \over 2} \right) \Gamma \left({n\epsilon \over 2}-1 \right) \Gamma\left(-{\epsilon \over 2} \right)}{\Gamma\left({(n-1)\epsilon \over 2}-1\right) \Gamma\left(2-{(n+1)\epsilon \over 2} \right)} \right\} \ ,
 \ee
 where $f(\epsilon)$ was introduced in (\ref{eq:fe}). The sum over $n$ in (\ref{sigma-pert-n})
 can be performed
 easily with the techniques introduced in \cite{pmp,pm} and summarized in
 Appendix \ref{pmp-trick}. The properties of the sum are encoded in the following structure function,
 \be
\label{struc-nlsm}
F(x,y) =- \left({p^2\over 4 \pi \nu^2}\right)^{-y/2}\frac{y}{x} \left(f (x) \right)^{y/x-1}\left\{ {\Gamma(2-y/2) \Gamma(y/2-1) \Gamma(1-x/2)\over \Gamma(y/2-x/2-1) \Gamma(2-y/2-x/2)} -f(x)\right\} \ .
\ee
Let us note that this function is the sum of two contributions. The first contribution is due to the diagrams
in \figref{bubble-fig} and to the first set of diagrams in \figref{more-bubble-fig}, while
the second contribution is due to the second set of diagrams in \figref{more-bubble-fig}. The whole structure function is finite at $x=0$, which is due to the cancellation of IR sub-singularities. In particular, we find,
%
\be
F(y)= F(0,y)=-\left(\frac{\mu}{p}\right)^y{y \over 2} \left( 2 \gamma_E + \psi \left( 2-{y \over 2} \right)+ \psi \left( {y \over 2}-1 \right) \right).
\ee
On the other hand, the scheme-dependent terms are governed by the function
\be
\label{f0-function}
F_0(x)=F(x,0)= -\frac{(x+2) \sin \left(\frac{\pi  x}{2}\right) \Gamma (-x)}{\pi  \Gamma \left(1-\frac{x}{2}\right) \Gamma
   \left(2-\frac{x}{2}\right)} \ .
   \ee
Therefore, one obtains the following representation of the self-energy :
\be
\ba
{1\over p^2} \Sigma_1^{\overline{\rm MS}}(p^2, \epsilon)&=1+\ln \frac{\lambda(p)}{\lambda(\mu)} +\int_0^{\infty}  \re^{- y/\lambda(p)}\biggl(-\gamma_E- {1\over 2} \psi \left( 2-{y \over 2} \right)-{1\over 2} \psi \left( {y \over 2}-1 \right) -{1\over y}\biggr)\rd y \\
& +F_0(\epsilon) \ln\left(1+\frac{\lambda(\mu)}{ \epsilon}\right)+ \CO(\epsilon) \ ,
 \label{eq:NLSMpert}
 \ea
 \ee
where we have used the RGE
\begin{align}
\left(\frac{\mu}{p}\right)^y \re^{-y/\lambda(\mu)}=\re^{- y/\lambda(p)} \ ,
\end{align}
to obtain the coupling constant at $\mu^2=p^2$.
The result (\ref{eq:NLSMpert})
was obtained in \cite{cr-dr} by studying the expression (\ref{sigma1}) in the limit
$m_0^2 \to 0$. It can be also obtained by using conventional perturbation theory around a vacuum in the NLSM which
breaks spontaneously the $O(N)$ symmetry \cite{sss,mmr-unpublished}.
It was noted in \cite{cr-dr} that (\ref{eq:NLSMpert}) allows us to calculate the renormalization
function $Z/Z_\lambda$ for the $\bPhi$ field, at NLO in $1/N$ but at all orders in $\lambda$. If we write the $1/N$ expansions
\be
Z= Z^{(0)}+ {1\over N} Z^{(1)}+ \cdots, \qquad Z_\lambda= Z_\lambda^{(0)}+{1\over N} Z_\lambda^{(1)}+ \cdots \ ,
\ee
we have
\be
\label{lead-ren}
Z^{(0)}= Z_\lambda^{(0)}= {1\over 1+ {\lambda \over \epsilon}} \ ,
\ee
so that
\be
\label{rf-field}
{Z \over Z_\lambda}=
1-{1\over N} \int_0^\lambda {\rd u \over u(u+ \epsilon)} \left(  {\beta^{(1)} (u) \over u} + \gamma^{(1)} (u)\right)+ \CO(N^{-2}) \ .
\ee
The term of order $1/N$ in (\ref{rf-field}) should cancel the divergence due to the second line in (\ref{eq:NLSMpert}). By using (\ref{F0div}) we conclude that the anomalous dimension
of the field in the $1/N$ expansion, in the $\overline{\text{MS}}$ scheme, is given by
\be
\gamma^{(1)} (\lambda)+{\beta^{(1)} (\lambda) \over \lambda}=\lambda F_0(-\lambda) \ .
\ee

The result in (\ref{eq:NLSMpert}) for the self-energy, which was obtained in the $\overline{\text{MS}}$ scheme,
is very similar to the result (\ref{Phi0}) obtained from the exact large $N$ solution in the SM scheme.
To make a more detailed comparison, we note that
the matching of schemes can be performed in the following way. The first observation is that the two-point functions
$S_{A,B}(p, g)$ in two different multiplicative schemes, $A$ and $B$, can be matched by using
\begin{align}
\label{match}
&\bigg(\lambda_A(\mu)\bigg)^{\frac{\gamma_0}{\beta_0}}
\exp\bigg[\int_0^{\lambda_A(\mu)} \left(\frac{\gamma_A}{\beta_A}(\lambda')-\frac{\gamma_0}{\beta_0 \lambda'}\right)\rd \lambda'\bigg]S_A(p,\lambda_A(\mu))\nonumber \\
&=c_{AB}\bigg(\lambda_B(\mu)\bigg)^{\frac{\gamma_0}{\beta_0}}\exp\bigg[\int_0^{\lambda_B(\mu)} \left(\frac{\gamma_B}{\beta_B}(\lambda')-\frac{\gamma_0}{\beta_0 \lambda'}\right)\rd \lambda'\bigg]S_B(p,\lambda_B(\mu))\ .
\end{align}
Here, $c_{AB}$ is a pure number and in particular it is $\mu$ independent. This is because the reversed RG
evolution factors in the two schemes compensate the dependence on $\mu$, and one is left
only with functions of $p^2$. Then, since this is essentially the same two-point function, the $\mu$-independent functions in the two schemes must only differ by an overall factor, which we call $c_{AB}$. It can be extracted by comparing the perturbative parts in the two schemes.

The ${\rm SM}$ scheme is a multiplicatively renormalizable scheme. It follows from (\ref{sigma1-ren})
that the two-point function at NLO in $1/N$ can be expressed in terms of the physical mass $m^2$ as
\begin{align}\label{2p-SM}
S_{\text{SM}}(p^2,M^2)=\frac{1-\frac{1}{N}\left(-\ln \lambda(M)/2+\gamma_E\right)+\CO(N^{-2}) }{p^2+m^2+\frac{1}{N}\Sigma_1^{\rm SM} (p)+\CO(N^{-2}) } \ .
\end{align}
The factor $-\ln \lambda(M)/2=\ln \ln (M^2/m^2)$ (this relation follows from the gap equation in the SM scheme, see \cite{biscari,cr-review})  is due to the field renormalization in this scheme. In fact, the anomalous dimension here is
\begin{align}
\frac{\beta_{\rm SM}(\lambda)}{\lambda}+\gamma_{\rm SM}(\lambda)=\frac{\lambda}{N}+{\cal O} \left(\frac{1}{N^2}\right) \ ,
\end{align}
which contains only a one-loop part, at this order in the $1/N$ expansion. It gives exactly the term $-\ln(\lambda(M))$ in the numerator. The scheme conversion factor $c_{AB}$ can then be found to be $1$. Indeed, the renormalized self-energy in the $\overline{\rm MS}$ scheme reads
\be
\ba
{1\over p^2} \Sigma_{1,R}^{\overline{\rm MS}}&=\left[ F_0(\epsilon) \ln \left(1+{\lambda(\mu)\over \epsilon} \right) \right]_{\rm fin}-\ln \lambda(\mu)\\
&+1+\ln \lambda(p)+ \int_0^{\infty}  \re^{- y/\lambda(p)}\biggl(-\gamma_E- {1\over 2} \psi \left( 2-{y \over 2} \right)-{1\over 2} \psi \left( {y \over 2}-1 \right)-{1\over y}\bigg)\rd y \ . \label{eq:MSrenorm}
\ea
\ee
The finite part in the first line can be extracted as
\begin{align}
\left[ F_0(\epsilon) \ln \left(1+{\lambda \over \epsilon} \right) \right]_{\rm fin}=\int_0^{\lambda} \frac{F_0(-u)-F_0(0)}{-u}  \rd u\ ,
\end{align}
which is exactly due to the regular part of the RG evolution, while the $-\ln \lambda(\mu)$ term is due to the singular part of the RG evolution. As such, in Eq.~(\ref{match}), after applying the reversed RG evolution, the first line of Eq.~(\ref{eq:MSrenorm}) is cancelled, leaving only the $\mu$ independent terms in the second line.  Similarly, in the Eq.~(\ref{2p-SM}), after applying the reversed RG evolution, the term $-\ln \lambda(M)$ in the numerator is cancelled, while the remaining $\ln 2+\gamma_E$ factor is also cancelled by that of the $\Phi_0(\lambda)$ given in Eq.~(\ref{Phi0}). In this way one ends up with identical $\mu^2,  M^2$ independent terms in the two schemes, thus $c_{AB}=1$. Equivalently, one can also extract $c_{AB}=1$ by setting $\mu^2=M^2=p^2$ and noticing that in this case the term $\ln \lambda(M)/2$ in the numerator of Eq.~(\ref{2p-SM}) cancels the term $\ln \lambda(p)/2$ of $\Phi_0(\lambda)$, while the factors $\lambda^{\gamma_0/\beta_0}$ in the two schemes cancel with each other.

\begin{figure}[!ht]
\leavevmode
\begin{center}
\includegraphics[height=6cm]{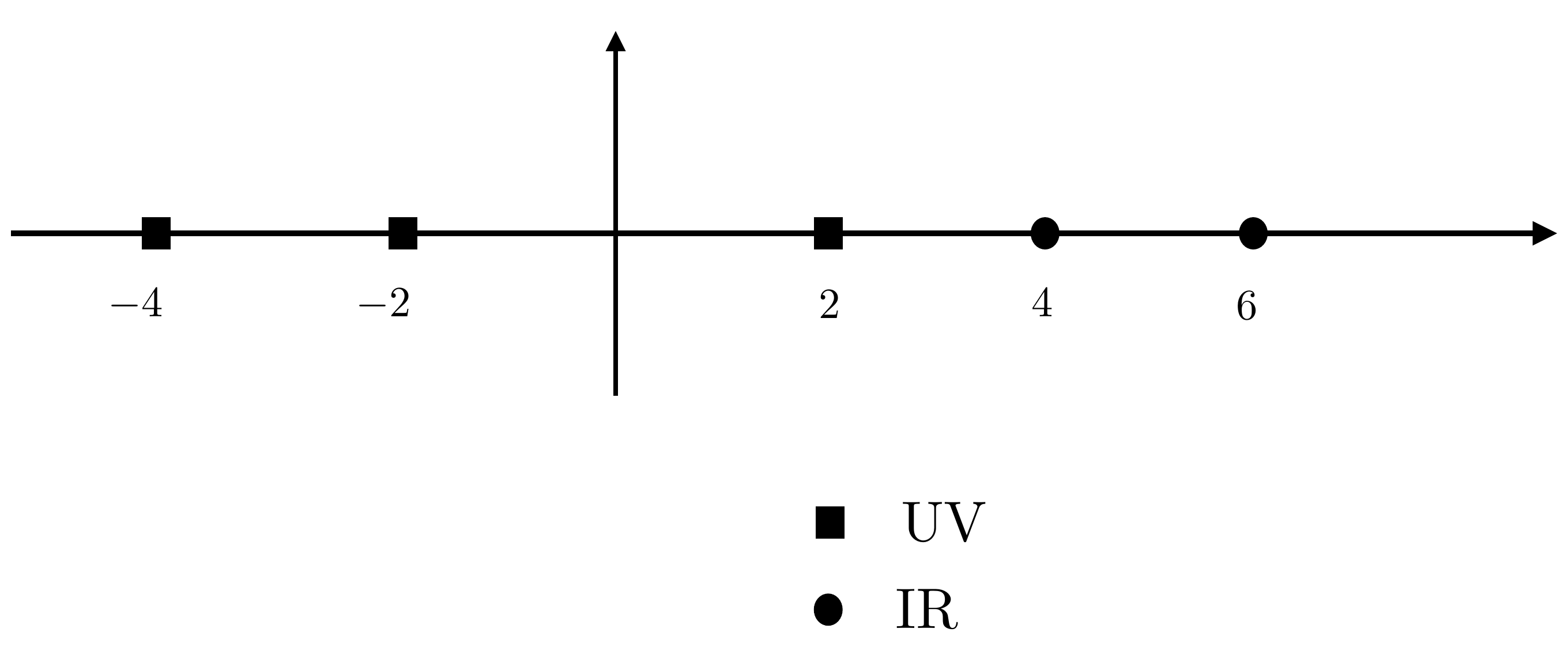}
\end{center}
\caption{Structure of renormalons for the perturbative self-energy Eq.~(\ref{eq:NLSMpert}).}
\label{fig:renormaself}
\end{figure}

Notice that the perturbative series for the self-energy Eq.~(\ref{eq:NLSMpert}) leads to singularities in the
Borel plane at $y\in 2 \mathbb{Z}$.
Let us review the standard argument to determine the nature of these singularities,
according to the momentum region in which they originate (see e.g. in \cite{beneke,grozin}). We first note that the renormalized polarization loop $\Pi_R(q^2)$ can be defined by renormalizing the coupling constant in the chain of bubbles inserted in (\ref{sigma-pert-n}), i.e.
\be
{1\over (2 \pi \lambda_0)^{-1} + q^2 \Pi(q^2, \epsilon)}= {2 \pi \lambda \over 1+ \lambda \Pi_R(q^2)}.
\ee
By using (\ref{lead-ren}) we find
\be
\Pi_R(q^2)= {1\over 2} \ln \left( {q^2 \over \mu^2} \right),
\ee
and the Borel resummation of the perturbative series in the the first line of (\ref{sigma-pert-n}) gives, up to an overall constant,
\be
\int {\rd^d q \over (2 \pi)^d} {q^2 \over (p-q)^2} \left( {q^2 \over \mu^2} \right)^{-y/2}.
\ee
In general, if the integrand of a resummed chain of bubbles behaves in the UV region $q^2 \to \infty$ as
\be
{1\over (q^2)^{n_{\rm UV}}} (q^2)^{-a y},
\ee
the singularities occurring at
\be
a y \le {d \over 2}- n_{\rm UV}
\ee
are UV renormalons. Similarly, if the integrand behaves in the IR region $q^2 \to 0$ as
\be
{1\over (q^2)^{n_{\rm IR}}} (q^2)^{-a y}
\ee
the singularities occuring at
\be
ay \ge {d \over 2}- n_{\rm IR}
\ee
are IR renormalons. In our case, $a=1/2$, $n_{\rm UV}=0$ and $n_{\rm IR} =-1$, so the singularities with $y\le 2$ are UV renormalons, while the singularities for $y\ge 4$ are IR renormalons, as shown in \figref{fig:renormaself}. In fact, the UV nature of the $y=2$ pole can be read from the arguments of the $\psi$ functions in Eq.~(\ref{eq:MSrenorm}). The $y=2$ singularity is only due to the term $\psi\left(\frac{y}{2}-1\right)$ in which the coefficient of $y$ is positive. According to the general rule stated below Eq.~(3.6) of~\cite{bb-polemass}, such poles are UV renormalons.

We conclude in particular that the first Borel singularity in the positive real axis, at $y=2$, is also an UV renormalon.
This renormalon is very special.
It appears because the perturbative self-energy is defined by Feynman integrals with
a superficial degree of UV divergence equal to $2$. This means that in cutoff regularizations there
will be power divergences proportional to the cutoff $\Lambda^2$, in addition to the finite part. If these power divergences are multiplied by divergent perturbative series with IR renormalon ambiguities, then after subtracting them out, the remaining finite parts will suffer from UV renormalon ambiguities, which is exactly the situation here. In the next subsection we will show that this UV renormalon is exactly cancelled by the ambiguity of a non-perturbative condensate, like
the usual situation for IR renormalons on the positive real axis.
However, this cancellation is not related to the ambiguity of separating UV and IR contributions in an OPE; it is rather a consequence of the absence of power divergences for the full two-point function, or equivalently, of multiplicative renormalizability. To our knowledge, this fact has not been discussed in detail in the literature. Although a similar UV renormalon has been observed and linked to power divergences in~\cite{jamin-ramon}, its cancellation with operator condensate ambiguities as a result of power divergence cancellation was not discussed therein. These points will be further elaborated in Sections \ref{sec:UVrenor} and~\ref{sec:renorUV}.

\subsection{Non-perturbative correction}\label{sec:OPEcond}

In the previous section we have shown that the perturbative self-energy in the NLSM can be obtained by considering the quartic linear sigma model
and taking the limit $\Lambda^2 \rightarrow \infty$. In this framework, we have to include an infinite number of condensates from the operator $\bPhi^2_0$. We now want to obtain the first non-perturbative correction in the NLSM from a condensate calculation in the linear sigma model.
We will verify that the result is in perfect agreement with the result (\ref{Phi1}) obtained in \cite{beneke-braun} from the exact large $N$ solution.

To obtain this contribution to the self-energy we have to include the following operators in the OPE for the two-point correlator
\begin{align}
&{\cal O}_{1,k}=-\ri D\left(\frac{1}{N}\bPhi_0^2\right)^k \ , \\
&{\cal O}_{2,k}=\frac{1}{N}\bPhi_0 \cdot \nabla^2  \bPhi_0 \times\left(\frac{1}{N}\bPhi_0^2\right)^k  \ .
\end{align}
Up to the desired $1/N$ order, only their tree-level condensates contribute, except for ${\cal O}_{1,0}$:
\begin{align}
&\langle {\cal O}_{1,0}\rangle=\langle X \rangle_0+\frac{\langle X \rangle_1}{N} \ ,\\
& \langle {\cal O}_{1,k}\rangle=\langle X \rangle_0 \left({1\over 2 \pi \lambda_0} \right)^k \ , \quad k\ge 1  \  ,\\
& \langle {\cal O}_{2,k} \rangle=\left\langle \frac{1}{N}\bPhi_0 \cdot \nabla^2  \bPhi_0 \right\rangle \left({1\over 2 \pi \lambda_0} \right)^{k} \ ,  \quad k\ge 0\ ,
\end{align}
where we have used factorization at large $N$, and we recall that the field $-\ri D$ becomes the field $X$ in the NLSM Lagrangian (\ref{lagPhi}).
The condensate $\langle X \rangle_1$ only attaches to the tree-level self-energy, while the tree-level condensates will be attached to the coefficient functions at order $1/N$.

\begin{figure}[!ht]
\leavevmode
\begin{center}
\includegraphics[height=3cm]{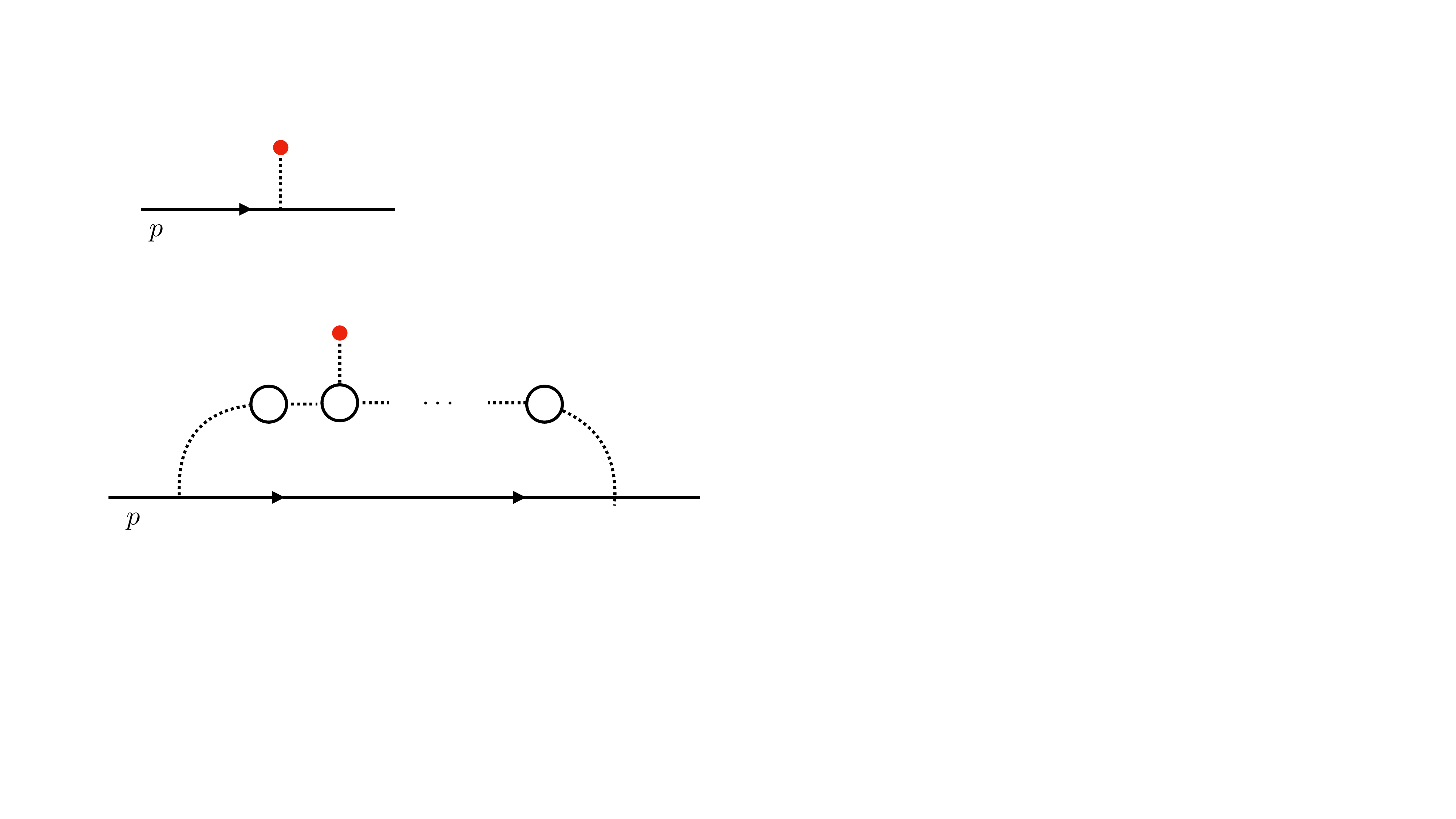}
\end{center}
\caption{The $X$ condensate attached to the $\bPhi_0$ propagator.}
\label{0D-fig}
\end{figure}

\begin{figure}[!ht]
\leavevmode
\begin{center}
\includegraphics[height=6.5cm]{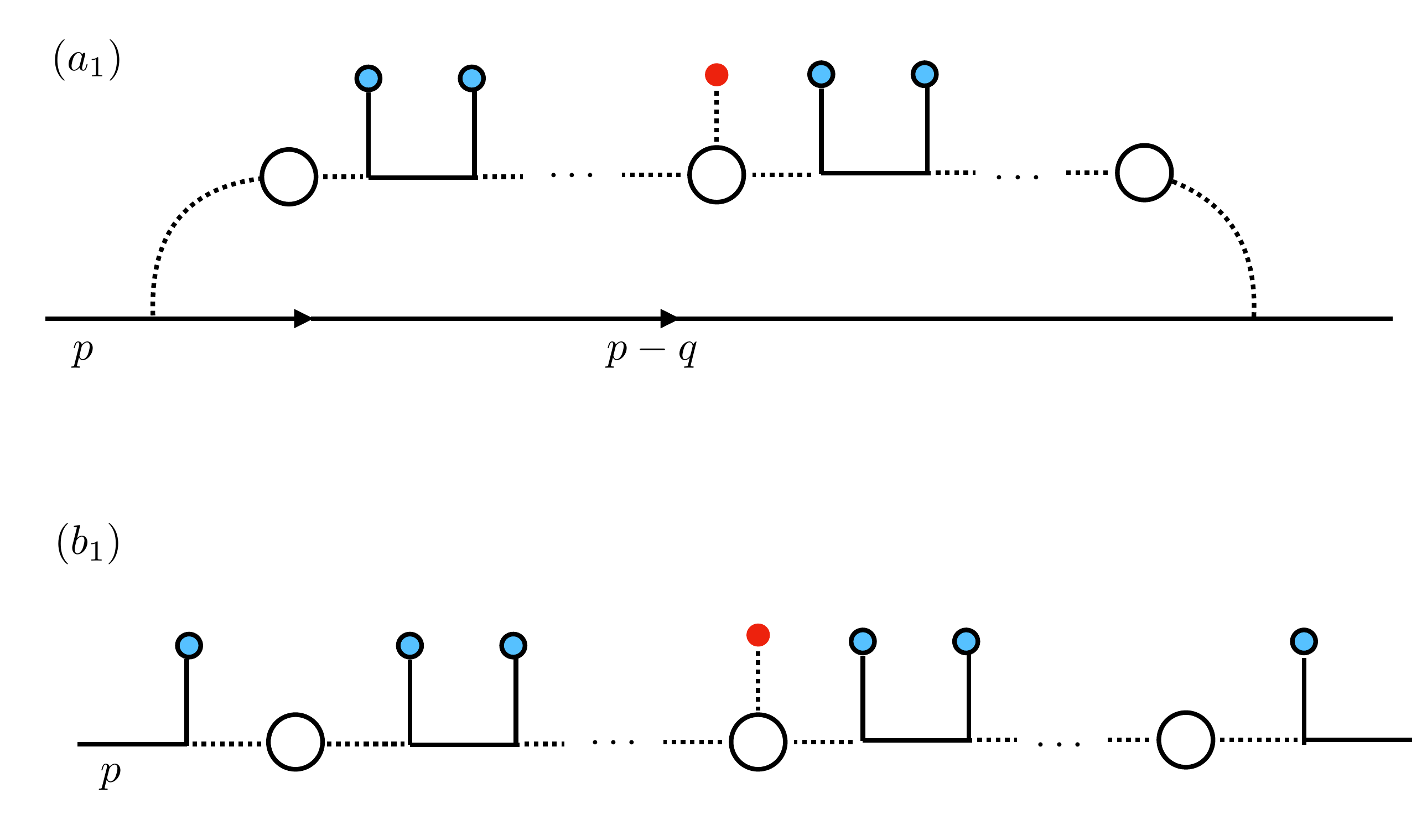}
\end{center}
\caption{The diagrams of type ($a_1$) and ($b_1$), obtained from the diagrams in \figref{more-bubble-fig} by
inserting one $\langle X\rangle_0$ condensate in a $\bPhi_0$ bubble.}
\label{a1b1-D-fig}
\end{figure}

Let us then compute these coefficient functions. For the operators ${\cal O}_{1,k}$, there are the
following types of contributions. The first contribution comes from $\CO_{1,0}$ and is given by the condensate
of the auxiliary field $X$ up to NLO in $1/N$, and attached to the
propagator of the $\bPhi_0$ field. This is shown in
diagram \figref{0D-fig} (condensates of the auxiliary field $X$, or equivalently $D$, will be
represented by a red blob). The diagrams that we will call of type ($a_1$) and ($b_1$) are obtained from the first class (respectively, second class) of diagrams in \figref{more-bubble-fig},
but inserting one single condensate of the $X$ field, $\langle X \rangle_0$, inside one of the bubbles of the $\bPhi_0$ field (we call it the special bubble). These diagrams are shown in \figref{a1b1-D-fig}.

\begin{figure}[!ht]
\leavevmode
\begin{center}
\includegraphics[height=6.5cm]{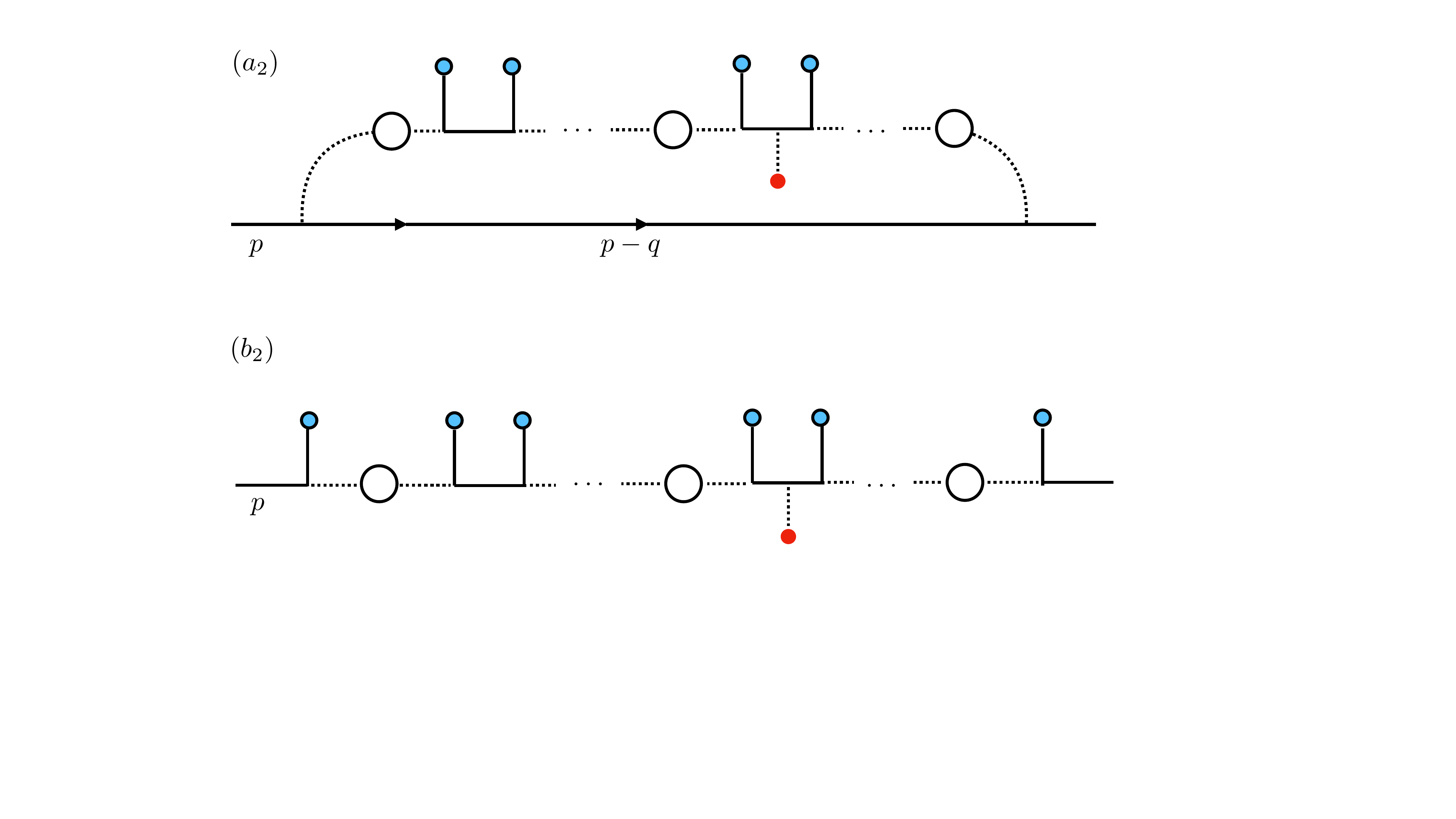}
\end{center}
\caption{The diagrams of type ($a_2$) and ($b_2$), obtained from the diagrams in \figref{more-bubble-fig} by
inserting one $\langle X\rangle_0$ condensate inside a $\bPhi_0$ propagator connecting two $\bPhi_0$ condensates.}
\label{a2b2-D-fig}
\end{figure}

\begin{figure}[!ht]
\leavevmode
\begin{center}
\includegraphics[height=3.5cm]{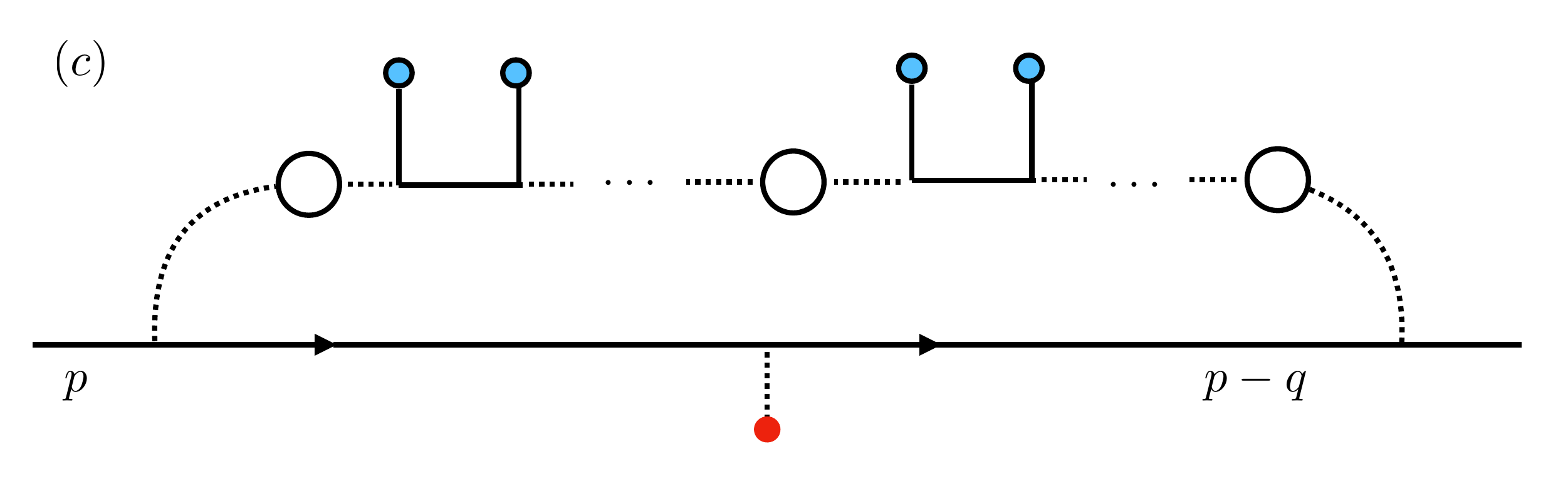}
\end{center}
\caption{The diagrams of type ($c$), obtained from the first diagram in \figref{more-bubble-fig} by
inserting one $\langle X\rangle_0$ condensate inside the $p-q$ line.}
\label{c-D-fig}
\end{figure}

In the diagrams that we will call of type ($a_2$), ($b_2$), we insert again one single $\langle X\rangle_0$ condensate in the diagrams of \figref{more-bubble-fig}, but this time inside the $\bPhi_0$ propagator connecting
two condensates of the $\bPhi_0$ field. They are depicted in \figref{a2b2-D-fig}. Finally, in the diagrams of
type ($c$), the condensate $\langle X\rangle_0$ is inserted on the $p-q$ line, while the
condensates of the $\bPhi_0$ field are in the bubble chain. See \figref{c-D-fig} for a depiction of these contributions.

The operators ${\cal O}_{2,k}$ appear in condensate calculus as follows. In the diagrams of \figref{more-bubble-fig}, the condensates of the $\bPhi_0$ fields are inserted at different points, and they have to be expanded as
\be
\label{cond-ex}
\langle \bPhi^a_0 (x) \bPhi^b_0 (0) \rangle = \langle \bPhi^a_0 (0) \bPhi^b_0 (0) \rangle +
{1\over 2 } x^{\mu} x^{\nu}  \langle \nabla_\mu \nabla_\nu \bPhi^a_0 (0)  \bPhi^b_0 (0) \rangle+ \cdots \ ,
\ee
(the condensate of the first derivative vanishes due to Euclidean invariance).
In the perturbative calculations in the previous section we only considered the first term in the r.h.s.
in (\ref{cond-ex}), but the derivative term contributes to the non-perturbative
correction that we are studying now (see e.g. \cite{pascual-tarrach} for an exposition of
this and other aspects of condensate calculus). The diagrams in \figref{more-bubble-fig} will lead to
three types of contributions due to the expansion of the condensate. The first diagram leads
to one contribution by expanding the condensate of a pair or $\bPhi_0$ fields
inserted in the $q$ line. The second diagram leads to two possible contributions: we can
expand the condensate of a pair or $\bPhi_0$ fields inside the line, or we can expand the
condensate of the pair of $\bPhi_0$ fields at the endpoints.


To calculate the contributions of all these diagrams, one needs the following functions:
\begin{align}
&\Pi_1(p^2,\epsilon)=-\int\frac{\rd^d q}{(2\pi)^d}\frac{1}{(p-q)^2 (q^2)^2}=-\frac{2}{p^2}(1+\epsilon)\Pi(p^2, \epsilon) \ , \\
&\Pi_2(p^2,\epsilon)=-\frac{1}{p^4} \ , \\
&\Pi_3(p^2,\epsilon)=\frac{1}{p^4}\frac{2+\epsilon}{2-\epsilon} \ .
\end{align}
The function $\Pi_1$ is the special bubble with one $\langle X \rangle$ condensate insertion. The function
$\Pi_2$ corresponds to the insertion of this condensate in the $\bPhi_0$ propagator joining the two $\bPhi_0$ condensates,
in \figref{a2b2-D-fig}. The function $\Pi_3$ gives the contribution of the operator $-\bPhi_0 \cdot \nabla^2 \bPhi_0$
in e.g. the  \figref{a2b2-D-fig}. To see this, we can evaluate the amputated condensate diagram
\be
\label{phi2cond}
\ba
\raisebox{-3.5ex}{\includegraphics[height=2cm]{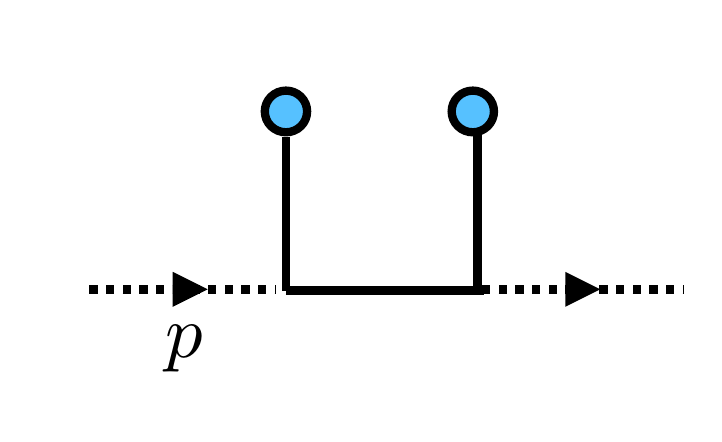}}&= \int  \re^{\ri p x} \wick{\c1{\bPhi^a_0} (x) \c1{\bPhi^b_0} (0) \langle \bPhi_0^a(x) \bPhi^b_0 (0) \rangle} \rd x \\
&= {\langle \bPhi^2_0 \rangle \over p^2} -{1\over p^4} {4-d \over d} \langle \bPhi_0 \cdot \nabla^2 \bPhi_0 \rangle+ \cdots \ .
\ea
\ee
In deriving this equation we simply inserted the Taylor expansion (\ref{cond-ex}), we integrated by parts, and we used that
\be
\langle  \bPhi_0 \cdot \nabla_\mu \nabla_\nu \bPhi_0 \rangle= {\delta_{\mu \nu} \over d}  \langle \bPhi_0 \cdot \nabla^2 \bPhi_0 \rangle \ ,
\ee
which follows from Euclidean invariance.

Let us first consider the coefficient functions associated to the diagrams in \figref{a1b1-D-fig} and \figref{a2b2-D-fig}. One can write
\be
\label{eq:CA11K}
\ba
 C_{1,\ell}^{(a_{1,2})}(p^2)&=\int\frac{\rd^d q}{(2\pi)^d}\frac{1}{(p-q)^2} {\cal G}_{1,\ell}^{(a_{1,2})}(q^2) \ ,  \\
 C_{1,\ell+1}^{(b_{1,2})}(p^2)&={\cal G}_{1,\ell}^{(a_{1,2})}(p^2) \ ,
 \ea
\ee
where $\ell$ is the number of insertions of $\bPhi_0^2$ condensates of the form (\ref{phi2cond}). As in the previous section,
the functions ${\cal G}_{1,\ell}^{(a_{1,2})}(q^2)$ are obtained by
summing over all bubbles and then taking the limit $\Lambda^2 \to \infty$. One finds
\be
\ba
{\cal G}_{1,\ell}^{(a_{1,2})}(q^2; \Lambda^2)&=\sum_{n=\ell+1}^{\infty}(-1)^{n}(\Lambda^2)^{n+1}n\binom{n-1}{\ell}\Pi_{1,2}\left(\frac{1}{q^2}\right)^\ell \Pi^{n-\ell-1} \\
&\rightarrow {\cal G}_{1,\ell}^{(a_{1,2})}(q^2)=-(-1)^\ell(\ell+1)\Pi_{1,2}\left(\frac{1}{q^2}\right)^\ell \Pi^{-\ell-2} \ .
\ea
\ee
The sum over the number of insertions can be performed in closed form and we obtain
\be \label{eq:sumgrand}
 \sum_{\ell=0}^{\infty}\bigg({\cal G}^{(a_1)}_{1,\ell}(q^2) \left({1\over 2 \pi \lambda_0} \right)^{\ell}+{\cal G}^{(a_2)}_{1,\ell}(q^2) \left({1\over 2 \pi \lambda_0} \right)^{\ell+1}\bigg) =2 \pi \lambda_0 {1 - 2 \pi \lambda_0 q^4 \Pi_1 \over \left(1+ 2 \pi \lambda_0 q^2 \Pi \right)^2}\ ,
\ee
which can be expanded in powers of $\lambda_0$. We then conclude that the contribution of the sets of diagram $(a_1)$, $(b_1)$, $(a_2)$ and $(b_2)$ is
\be
2 \pi \lambda_0 \int\frac{\rd^d q}{(2\pi)^d}\frac{1}{(p-q)^2}{1 - 2 \pi \lambda_0 q^4 \Pi_1 (q^2,\epsilon) \over \left(1+ 2 \pi \lambda_0 q^2 \Pi (q^2, \epsilon) \right)^2}+ {1 - 2 \pi \lambda_0 p^4 \Pi_1 (p^2,\epsilon) \over \left(1+ 2 \pi \lambda_0 p^2 \Pi  (p^2,\epsilon)\right)^2} \ .
\ee
The contribution of the diagrams of type $(c)$ is easy to obtain since the insertion of the condensate in the $p-q$ line just leads to an insertion of an extra propagator in (\ref{bub+cond}). It is given by
\be
-\int \frac{\rd^d q}{(2\pi)^d}\frac{q^2 }{(p-q)^4}\frac{2 \pi \lambda_0 }{1+2 \pi \lambda_0 q^2 \Pi(q^2,\epsilon)} \ .
\ee
In the equations above we have not included explicitly the factors of $N$ to simplify the presentation. It is easy to see that in the end one obtains
an overall factor of $1/N$, as in the previous section.

We now expand in powers of $\lambda_0$, and perform the integrals explicitly with standard one-loop formulae.  We find the following overall contribution of the $\langle X \rangle_0$ condensate
\be
 1+\sum_{n \ge 1} \left(D_n^{(a_1)} + D_n^{(b_1)} + D_n^{(a_2)}+D_n^{(b_2)}+ D_n^{(c)} \right) \lambda_0^n \ ,
\ee
where
\be
\ba
D_n^{(a_1)}&=-(1+ \epsilon)  (n-1) \left( \frac{p^2}{4\pi} \right)^{-n\epsilon/2} \left( {f(\epsilon) \over \epsilon} \right)^{n-1} \frac{\Gamma(-\frac{\epsilon}{2}) \Gamma\left(\frac{n\epsilon}{2}\right) \Gamma\left(1-\frac{n\epsilon}{2}\right)}{\Gamma\left(\frac{(n-1)\epsilon}{2}\right) \Gamma\left(1 - \frac{(n+1)\epsilon}{2}\right)} \ ,\\
D_n^{(b_1)}&=-2 (1+ \epsilon)  n \left( \frac{p^2}{4\pi} \right)^{-n\epsilon/2}  \left( {f(\epsilon) \over \epsilon} \right)^{n} \ ,\\
D_n^{(a_2)}&={n   \over 2} \left( \frac{p^2}{4\pi} \right)^{-n\epsilon/2} \left( {f(\epsilon) \over \epsilon} \right)^{n-1} \frac{\Gamma(-\frac{\epsilon}{2}) \Gamma\left(\frac{n\epsilon}{2}\right) \Gamma\left(1-\frac{n\epsilon}{2}\right)}{\Gamma\left(\frac{(n-1)\epsilon}{2}\right) \Gamma\left(1 - \frac{(n+1)\epsilon}{2}\right)} \ ,\\
D_n^{(b_2)}&= (n+1) \left( \frac{p^2}{4\pi} \right)^{-n\epsilon/2} \left( {f(\epsilon) \over \epsilon} \right)^{n}  \ ,\\
D_n^{(c)}&=-{1\over 2} \left( \frac{p^2}{4\pi} \right)^{-n\epsilon/2}  \left( {f(\epsilon) \over \epsilon} \right)^{n-1}  \frac{\Gamma(2-\frac{n\epsilon}{2}) \Gamma\left(\frac{n\epsilon}{2}\right) \Gamma\left(-1-\frac{\epsilon}{2}\right)}{\Gamma\left(\frac{(n-1)\epsilon}{2}-1\right) \Gamma\left(1 - \frac{(n+1)\epsilon}{2}\right)} \ .
\ea
\ee
Notice that the function $f(\epsilon)$ was defined in (\ref{eq:fe}).

Let us now consider the contributions of the condensate of $\bPhi_0 \cdot \nabla^2 \bPhi_0/N$,
or equivalently, the coefficient functions of the
operators $\CO_{2,k}$ in the OPE. As we mentioned above, there are
three types of contributions. Two of these come from considering the diagrams in \figref{more-bubble-fig}
and expanding one pair of condensates as in (\ref{phi2cond}). Their contributions will be
denoted $(a)$, $(b)$, referring respectively to the first and second diagrams in \figref{more-bubble-fig}.
They can be obtained as in the previous calculation:
\be
\label{eq:CA22K}
\ba
 C_{2,\ell}^{(a)}(p^2)&=\int\frac{\rd^d q}{(2\pi)^d}\frac{1}{(p-q)^2} {\cal G}_{2,\ell}^{(a)}(q^2) \ ,  \\
 C_{2,\ell+1}^{(b)}(p^2)&={\cal G}_{2,\ell}^{(a)}(p^2) \ ,
 \ea
\ee
where
\be
{\cal G}_{2,\ell}^{(a)}(q^2)=(-1)^\ell(\ell+1)\Pi_{3}\left(\frac{1}{q^2}\right)^\ell \Pi^{-\ell-2} \ .
\ee
The third contribution, which will be denoted by $(c)$, is slightly different, since the condensates are
inserted at the endpoints of the propagator. When integrating by parts one
obtains the operator
\be
\label{oper-eq}
\frac{\delta_{\alpha\beta}}{2d}\frac{\partial^2 }{\partial p_\alpha \partial p_\beta}=
{\partial \over \partial p^2} + {2 p^2 \over d} {\partial^2 \over \partial (p^2)^2} \ ,
\ee
acting on the whole chain (\ref{chain-sum}) (we note that (\ref{oper-eq}) is of course only
valid when both sides act on functions of $p^2$). Collecting
everything, we find
\be
\ba
& 2 \pi \lambda_0 {4-d\over d} \left\{  \int\frac{\rd^d q}{(2\pi)^d}\frac{1}{(p-q)^2}
{2 \pi \lambda_0 \over (1+ 2 \pi \lambda_0 q^2 \Pi(q^2, \epsilon))^2}+
{1\over (1+ 2 \pi \lambda_0 p^2 \Pi(p^2, \epsilon))^2} \right\} \\
& - 2 \pi \lambda_0 \left({\partial \over \partial p^2} + {2 p^2 \over d} {\partial^2 \over \partial (p^2)^2}\right)
 {p^2 \over 1+ 2 \pi \lambda_0 p^2 \Pi(p^2, \epsilon)} \ .
\ea
\ee
Like before, we can expand in $\lambda_0$ and perform the integrals. The final result is given by
\be
2 \pi \lambda_0 \left\{ {2 \epsilon  \over 2-\epsilon}+  \sum_{n \ge 1} \left(E_n^{(a)} + E_n^{(b)} + E_n^{(c)} \right) \lambda_0^n \right\} \ ,
\ee
where
\be
\ba
E_n^{(a)}&={n\over 2} \frac{2+\epsilon}{2-\epsilon} \left( \frac{p^2}{4\pi} \right)^{-n\epsilon/2} \left( {f(\epsilon) \over \epsilon} \right)^{n-1} \frac{\Gamma(-\frac{\epsilon}{2}) \Gamma\left(\frac{n\epsilon}{2}\right) \Gamma\left(1-\frac{n\epsilon}{2}\right)}{\Gamma\left(\frac{(n-1)\epsilon}{2}\right) \Gamma\left(1 - \frac{(n+1)\epsilon}{2}\right)},\\
E_n^{(b)}&=(n+1) \frac{2+\epsilon}{2-\epsilon}  \left( \frac{p^2}{4\pi} \right)^{- n\epsilon/2} \left( {f(\epsilon) \over \epsilon} \right)^{n},\\
E_n^{(c)}&= -\frac{(2-(n+1)\epsilon)(2-n\epsilon)}{2(2-\epsilon)} \left( \frac{p^2}{4\pi} \right)^{-n\epsilon/2}  \left( {f(\epsilon) \over \epsilon} \right)^{n} \ .
\ea
\ee
We now note that, due to the EOM (\ref{eom}),
\be
2 \pi \lambda_0 \left\langle {1\over N} \bPhi_0 \cdot \nabla^2 \bPhi_0\right\rangle= \langle X \rangle_0  \ ,
\ee
therefore we should add the contributions $D_n^{(\cdot)}$, $E_n^{(\cdot)}$. To do this, we can use again the
formalism of Appendix \ref{pmp-trick}. We introduce the following functions:
\be
\ba
\mathfrak{b}_1(x,y)&= \frac{\Gamma \left(-\frac{x}{2}\right) \Gamma \left(1-\frac{y}{2}\right) \Gamma
   \left(\frac{y}{2}\right)}{\Gamma \left(\frac{1}{2} (-x-y)+1\right) \Gamma
   \left(\frac{y-x}{2}\right)} \ , \\
   \mathfrak{b}_2(x,y)&=\frac{\Gamma \left(-\frac{x}{2}-1\right) \Gamma \left(2-\frac{y}{2}\right) \Gamma
   \left(\frac{y}{2}\right)}{\Gamma \left(\frac{1}{2} (-x-y)+1\right) \Gamma
   \left(\frac{y-x}{2}-1\right)}\ .
   \ea
   \ee
   The total structure function for this problem is,
\be
\ba
N(x, y)= \left( {p^2 \over \re^{\gamma_E}\mu^2} \right)^{-y/2}  \left(f(x)\right)^{y/x-1}& \biggl\{
\frac{ y \left(x^2-x (y+1)+y-2\right)}{x-2}\mathfrak{b}_1(x,y)-{y\over 2} \mathfrak{b}_2(x,y)\\
& \, \, + \frac{ y
   \left(-x (3 y+2)+y^2-4\right)}{2 (x-2) x}f(x) \biggr\} \ ,
   \ea
   \ee
 This function turns out to be regular at $x=y=0$. As in the perturbative result, many of the individual coefficients lead to a singular result when $x \to 0$, and it is only when summing all diagrams that we find a regular result. This is indeed due to the cancelation of IR singularities. The relevant information in this function are the values $N_0(x)=N(x,0)$
 and $N(y)= N(0,y)$. We find, for $p^2= \mu^2$,
 \be
 \label{stru-ny}
 \ba
 N(y)&=  {y\over 2}\left( {y^2 \over 4}-1 \right) \left( 2 \gamma_E + \psi \left( 2-{y \over 2} \right) + \psi \left(1+{y \over 2} \right) \right)+1-{y\over 2} +{y^2 \over 2}-{y^3 \over 4}\\
 &=1-y + {y^2 \over 4} +{y^3 \over 4} (\zeta(3)-1) + \cdots \ ,
 \ea
 \ee
 We can now compare this result for the first exponential correction
 to the one obtained with the exact large $N$ method in (\ref{n1fgh}) and (\ref{Phi1}).
 It is easy to see that (\ref{stru-ny}) leads precisely to the formal power series appearing in (\ref{Phi1}. The function
 \be
 N_0(x)= -\frac{2 (2 x+1) \sin \left(\frac{\pi  x}{2}\right) \Gamma (-x)}{\pi  \Gamma
   \left(1-\frac{x}{2}\right)^2} \ ,
   \ee
determines the divergences.

 The non-perturbative self-energy, as calculated from the condensates, has two pieces. The first piece is given by the condensate $\langle X\rangle$, due to the condensate diagram in \figref{0D-fig}. The second piece, which we will denote by
 $\Sigma_{1, \text{NP}}^{\overline{\text{MS}}}(p^2, \epsilon)$, can be read from the above results as
\be
\label{fin-result}
{1\over\langle X \rangle_0} \Sigma_{1, \text{NP}}^{\overline{\text{MS}}}(p^2, \epsilon)=N_0(\epsilon) \ln\left(1+ {\lambda \over\epsilon} \right)
+1+ \int_0^\infty \re^{-y/ \lambda} {N(y) -1 \over y} \rd y+ \CO(\epsilon) \ .
\ee
The first term in the r.h.s. contains the divergence
 \be
 \left[ N_0(\epsilon) \ln \left(1+ {\lambda \over \epsilon} \right) \right]_{\rm div}= \int_0^\lambda {N_0(-u) \over u+ \epsilon} \rd u \ ,
 \ee
 and has to be renormalized.  The renormalization of the self-energy requires multiplying by (\ref{rf-field}),
 while the renormalization of the operator in the condensate gives
 \be
 \langle X \rangle= \langle [X] \rangle Z_\lambda \left(1- {\beta_\lambda \over \epsilon \lambda} \right)= \langle [X] \rangle \left\{1+{1\over N}\int_0^\lambda {\rho(u)
 \over u (u+\epsilon)} \rd u + \CO(N^{-2})  \right\} \ .
 \ee
This is due to (\ref{ren-op}) and the EOM (\ref{eom}). Since
 \be
 \label{N0}
 N_0(-u)= -F_0(-u)- {\rho(u) \over u} \ ,
 \ee
all divergences cancel, as it should. In fact, the condensate/OPE calculation performed in this section
could be used to determine the $1/N$ correction to the beta function
of the NLSM (a different diagrammatic calculation can be found in \cite{mmr-testing}). Let us note that the finite part of the first term in
the r.h.s. of (\ref{fin-result}) is also due to the regular part of the RG evolution, as in the perturbative result, while the
additional factor $\ln(\lambda/2)$ in (\ref{Phi1}) is due to the singular part.

As in \cite{tsc}, one can compare the exact large $N$ calculation of (\ref{Phi1}) to the above condensate calculation, in order to obtain the
condensate $\langle X \rangle_1$ at NLO in $1/N$. Indeed, if one knows the coefficient function
for the operator $\CO$ in a scheme $A$, the full two-point function in the scheme $B$, and the scheme conversion factor
$c_{AB}$ in (\ref{match}), then one can extract the non-perturbative value for the operator condensate as
\begin{align}
c_{\CO,A}=\bigg(\lambda_A(\mu)\bigg)^{\frac{\gamma_{\CO,0}}{\beta_0}}\exp\bigg[\int_0^{\lambda_A(\mu)}\rd g' \left(\frac{\gamma_{\CO,A}}{\beta_A}(\lambda')-\frac{\gamma_{\CO,0}}{\beta_0 \lambda'}\right)\bigg]\langle \Omega |\CO_A|\Omega\rangle (\lambda_A(\mu))\ ,
\end{align}
in the scheme $A$ (here, $|\Omega\rangle$ denotes the physical vacuum with non-trivial condensates). This is due to
the relation
\begin{align}
&\bigg(\lambda_A(\mu)\bigg)^{\frac{\gamma_0}{\beta_0}}\exp\bigg[\int_0^{\lambda_A(\mu)}\rd g' \left(\frac{\gamma_A}{\beta_A}(\lambda')-\frac{\gamma_0}{\beta_0 \lambda'}\right)\bigg]C_A(p^2,\lambda_A(\mu))\CO_A(\lambda_A(\mu)) \nonumber \\
&=c_{AB}\bigg(\lambda_B(\mu)\bigg)^{\frac{\gamma_0}{\beta_0}}\exp\bigg[\int_0^{\lambda_B(\mu)}\rd '\lambda' \left(\frac{\gamma_B}{\beta_B}(\lambda')-\frac{\gamma_0}{\beta_0 \lambda'}\right)\bigg]C_B(p^2,\lambda_B(\mu))\CO_B(\lambda_B(\mu)) \ .
\end{align}
If one knows $C_B(p^2,\lambda_B(\mu))\CO_B(\lambda_B(\mu))$, which is case if the full two-point
function in the scheme $B$ is known and $\CO$ is the only operator contribution at a
given power, then the above two relations can be combined to determine $c_{\CO,A}$. In the
previous section we determined $c_{AB}$ by comparing the perturbative part. By comparing the
$\overline{\rm MS}$ coefficient function for $X$ with the full SM result at the next-to-leading power,
one can extract also the non-perturbative number for $X$ in the $\overline{\rm MS}$ scheme.

Let us then compare the non-perturbative corrections in the different schemes.
The $-\gamma_E$ in (\ref{Phi1}) is again cancelled by the numerator in (\ref{2p-SM}).
In this way, after taking into account the $m^2$ at LO, one can extract
the $\mu$-independent part of the $X$ condensate in the $\overline{\rm MS}$ scheme as
\be
\label{cond-value}
\exp\bigg[\int_0^{\lambda(\mu)}\frac{\gamma_{X}(u)}{\beta(u)}\rd u \bigg]\langle \Omega| X|\Omega\rangle(\lambda(\mu))=m^2\left(1 \pm \frac{\ri\pi}{N}+{\cal O}\left(\frac{1}{N^2}\right)\right)=c_X(N)\Lambda^2_{\overline{\rm MS}} \ .
\ee
The $\pm \ri\pi$ appearing here is correlated with the renormalon ambiguity at $y=2$.
Notice that, since $\gamma_X$ starts at two loops, its
evolution is regular. Thus the constant for the condensate is free from $\ln 2$ ambiguities due to the
different normalization of the coupling. In (\ref{cond-value}), $m$ is the physical mass of the NLSM up to NLO in $1/N$. After converting it to the $\Lambda_{\overline{\rm MS}}$ parameter \cite{hmn,hn}, one has
\begin{align}
c_X(N)=1+\frac{1}{N}\bigg(6\ln 2+2\gamma_E-2\pm \ri\pi \bigg)+{\cal O}\left(N^{-2}\right) \ .
\end{align}
This is very similar to the result for the four-quark condensate in the Gross-Neveu
model obtained in \cite{tsc}, as the real parts are all equal to the physical mass squared, while
the strength of the renormalon differs a factor of two. In fact, by
multiplying the result above by a factor of $\frac{\beta}{2g^2}$, one obtains the following average of the trace anomaly 
\begin{align}
\left\langle \Omega \left |-\frac{\beta(g)}{2g^2}(\partial \bss)^2\right|\Omega \right\rangle=-\frac{m^2}{4\pi} \left(N-2\pm \ri\pi+{\cal O} \left(\frac{1}{N}\right)\right)=-m\frac{\partial F(m)}{\partial m}\ ,
\end{align}
which is consistent with a direct computation of the free
energy $F(m)$ in \cite{biscari,cr-review}.
In addition, the ambiguous imaginary part agrees with the result of
\cite{mmr-testing,mmr-an}\footnote{This imaginary ambiguity in the free energy of the NLSM is correlated with an UV renormalon which was discovered in \cite{mmr-testing} and further studied in \cite{mmr-an}. However, in those papers it was incorrectly labeled as an IR renormalon.}. This condensate is the analogue of the gluon
condensate $\frac{\beta}{2g}\langle F^2\rangle$ in massless QCD.

As a final application of our results we can also calculate the propagator of the $X$ field and its
first non-perturbative correction. The perturbative
part is simply given by the second diagram of \figref{more-bubble-fig}, i.e. by
\be
\label{pixp}
\Pi_{X}^{\rm P}(p^2, \epsilon)= {1\over N} {2 \pi \lambda_0 p^2\over  1+2 \pi \lambda_0 p^2 \Pi(p^2, \epsilon)}= {1\over N}  {4 \pi p^2 \over \ln\left(p^2/\Lambda_{\overline{\text{MS}}}^2 \right)}+ \CO(\epsilon) \ .
\ee
The non-perturbative correction is given by the sum of the diagram $(b_1)$ in
\figref{a1b1-D-fig}, the diagram $(b_2)$ in \figref{a2b2-D-fig}, and the
contribution of the second diagram of \figref{more-bubble-fig}, after expanding the
condensate as in (\ref{phi2cond}). One finds,
\begin{align}
\Pi_{X}^{\rm NP}(p^2, \epsilon)&={1\over N} \langle X \rangle \sum_{n \ge 0} (-2  \pi p^2 \Pi(p^2, \epsilon))^n \lambda_0^{n+1} \left\{ {4 -d \over d}(n+1)+ n+1 -2 (1+ \epsilon) n \right\} \ .
\end{align}
This can be summed up and is finite in the $\epsilon \to 0$ limit; it gives
\be
\label{pixnp}
\Pi_{X}^{\rm NP}(p^2)= {8 \pi m^2 \over N}  \left( {1 \over  \ln\left(p^2/\Lambda_{\overline{\text{MS}}}^2 \right)}- {1 \over  \ln^2 \left(p^2/\Lambda_{\overline{\text{MS}}}^2 \right)} \right)\, ,
\ee
which, at this order in $1/N$, is precisely the expansion at large $p^2/m^2$ of the exact propagator (\ref{propa-X}).
The contribution of the leading inverse logarithm in (\ref{pixnp}) was obtained in section 3.2 of \cite{svz-pr} from a
condensate calculation, albeit with considerable effort. With our method, the whole non-perturbative correction follows from
a simple diagrammatic approach.

\subsection{Power divergences and the UV renormalon}
\label{sec:UVrenor}

Let us consider more carefully the nature of the UV renormalon at $y=2$. We have seen in the previous section
that it cancels against the ambiguity of the $X$ condensate. As we mentioned at the end of Sec. \ref{sec:OPEpert}, this
unusual cancellation is the consequence of two different, more conventional
cancellations: the UV renormalon of the perturbative series cancels against a renormalon multiplying the power divergence of the self-energy diagram, while the ambiguity in the $X$ condensate cancels against a renormalon multiplying the power divergence of the tadpole diagram.
However, multiplicative renormalizability of the theory implies that these power divergences cancel among themselves, and this explains
the unusual cancellation observed in the previous section.

To see this explicitly, we have to choose a regularization scheme where power divergences are manifest. One possibility is the SM scheme we used in section \ref{sec-rev}, since it depends on the scale $M$ in addition to $m_0^2$. We have already calculated the contribution of the tadpole diagram and the self-energy diagram in that scheme, in the exact large $N$ method, and we repeat it here for the reader's convenience:
\be
\label{smt-2}
\Sigma^1_T(m_0^2, M^2)=M^2\int_0^{\infty}  \frac{1}{y-2}\left(\frac{M}{m_0}\right)^{-y}\rd y+\widehat m_1^2,
\ee
The first term in the r.h.s. is the power divergence. Its pole cancels against the ambiguity in $\widehat m_1^2$,
which is the $1/N$ contribution to the condensate $\langle X \rangle$.

Let us now consider the self-energy diagram. We are interested in its perturbative limit, which contains the renormalon at $y=2$. This is easily obtained from (\ref{ssreg-t}) by taking the limit $m_0^2 \rightarrow 0$. This means that, in the renormalized $1/N$ correction to the self--energy (\ref{sigmaSM}), which has a trans-series representation, we keep only the perturbative part. In the end we obtain,
\be
\label{sms}
\Sigma_{S, \rm{p}}^1(p, m_0^2, M^2)= -M^2 \int_0^{\infty}  \frac{1}{y-2}\left(\frac{M}{m_0}\right)^{-y}\rd y + p^2 \left( \gamma_E + \ln \ln \left( {M^2 \over m_0^2} \right) + \Phi_0(\lambda(p)) \right),
\ee
where we have kept as wlel the the power divergence proportional to $M^2$. The pole at $y=2$ in this power divergence cancels now against the $y=2$ singularity of $\Phi_0$\footnote{Let us note that, as compared with the original expression (\ref{ssreg-t}), in taking the perturbative limit we have removed the contribution of the second term, but also the contribution of $H_1(y)$ in Eq.~(\ref{n1fgh}), which cancels the $y=2$ pole of $\Phi_0$. The net effect is that, as stated, the renormalon pole in the self-energy cancels against the power divergence in the same quantity.}. This further supports the statement that the
$y=2$ renormalon of the perturbative self-energy is an UV renormalon. If we now add (\ref{smt-2})
and (\ref{sms}), the power divergences exactly cancel, as they
should. The overall result is still free from ambiguities, but now the UV renormalon of the
power series $\Phi_0$ cancels against the ambiguity in $\widehat m_1^2$, which is the ambiguity in the
condensate of $X$. As such, we conclude that this unusual cancellation is a consequence of the power-divergence cancellation between the two diagrams.


\section{More on the relationship between the LSM and the NLSM}
\label{sec:LSM}

In the previous section, the NLSM has been represented as a limit of the LSM at the {\it integrand level}. This
allows to decouple the nonphysical
scale $\Lambda$ and to perform the OPE computation directly for the NSLM, without
worrying about UV effects belonging to the quartic model. One might ask what happens if one
reverses the order of taking the limits, and studies the weak-coupling limit of the LSM
as a two dimensional QFT. In this case, one might worry that $\Lambda$ becomes a
dynamical scale with non-trivial physics that does not decouple from the NLSM in the IR. In this section we show that,
at least to NLO in the $1/N$ expansion, this does not happen: there does exist non-trivial physics at the
cutoff scale, but it factorizes from the NLSM in the IR.

To present our results, let us consider the quartic LSM with a negative squared mass given by
\be
\label{lsm-action}
S=\int {\rm d}^2x \bigg(\frac{1}{2}(\nabla \bPhi_0)^2-\frac{M^2}{4}\bPhi_0^2+\frac{\pi M^2\gamma}{2N}{\cal N}_{M^2}(\bPhi_0^2)^2\bigg) \ .
\ee
It relates to the representation in Eq.~(\ref{eq:deflinear}) through
$\lambda_0=2\gamma$ and $\Lambda^2=\gamma M^2/N$. As a super-renormalizable
QFT, it requires only a mass renormalization for correlation functions. We perform the mass
renormalization by using the normal ordering of the quartic term at the mass scale $M^2$ (see e.g. \cite{slava})
\begin{align}
&{\cal N}_{M^2}(\bPhi_0^2)^2=(\bPhi_0^2)^2-2(N+2){\cal I}(M^2)\bPhi_0^2+N(N+2){\cal I}^2(M^2) \ , \\
&{\cal I}(M^2)=\int \frac{{\rm d}^dk}{(2\pi)^d}\frac{1}{k^2+M^2} \ .
\end{align}
$M^2$ is the naive mass for the ``Higgs particle'' that appears when we expand around the classical minima
of the potential which break the $O(N)$ symmetry.

Due to the ambiguity in the normal ordering scale, the division between
the quartic and the quadratic terms is not strict. We can equivalently represent the
theory with a positive squared mass term without changing the correlation functions of the
$\bPhi_0$ field, in a way similar to the duality relation for the one-component theory~\cite{Chang:1976ek}:
\begin{align}\label{def:quartic}
S=\int {\rm d}^2x \bigg(\frac{1}{2}(\nabla \bPhi_0)^2+\frac{m_0^2}{2}\bPhi_0^2+\frac{\pi m_0^2 \hat g }{2N}{\cal N}_{m_0^2}(\bPhi_0^2)^2\bigg) \ .
\end{align}
Here, the normal ordering is performed at a different mass $m_0^2$. The two
dimensionless couplings and the masses relate to each other through
\begin{align}
m_0^2=M^2\frac{\gamma}{\hat g}, \qquad  \frac{1}{\gamma}+\left(1+\frac{2}{N}\right)\ln \frac{\gamma}{\hat g}=-\frac{2}{\hat g} \ .
\end{align}
In particular, the weak coupling limit $\gamma \rightarrow 0$ of the
LSM (\ref{lsm-action}) is mapped to the strong coupling limit $\hat g \rightarrow \infty$ of the
positive mass squared version. In the $\gamma \rightarrow 0$ limit,
the mass $m_0^2$ is {\it exponentially small} in the coupling constant $\gamma$~\cite{Sberveglieri:2020eko}
\begin{align}\label{eq:massnormal}
m_0^2=M^2\exp \bigg[-\left(1+\frac{2}{N}\right)^{-1}\frac{1}{\gamma}\bigg]\left(1+{\cal O}\left({\rm e}^{-\frac{1}{\gamma}}\right)\right) \ .
\end{align}
This relation is already very similar to the mass relation in the NLSM and suggests that the theory becomes the NLSM around the scale $m_0^2$. However, the $M^2$ is another scale of the theory and plays the role of an UV cutoff, at which there are still UV fluctuations. The best one can hope is that these UV fluctuations factorize from the NLSM. In the next section we will show that this is indeed the case.

\subsection{Perturbative self-energy from the LSM}\label{sec:LSMpert}
Eq.~(\ref{eq:massnormal}) suggests that the LSM reduces to the NLSM at the exponentially small scale $m_0^2\sim M^2 \re^{-1/\gamma} \ll M^2$ when $\gamma \rightarrow 0$. This means, in the physical units of the NLSM, that to decouple the UV physics one needs to take the limit $M^2 \rightarrow \infty$, which is consistent with the limit $\Lambda^2 \rightarrow \infty$ adopted in the previous section.  Equivalently, one must send the momentum to zero, in units of $M$. As such, the first step to see the connection between the two models is to show that the perturbative framework for the LSM reduces to that of the NLSM in the small momentum limit.

Indeed, since the LSM (\ref{lsm-action}) is still weakly coupled at the scale $M^2$, one expects that there should be a perturbative way to compute the $\gamma^n$ corrections to the correlation functions, if all the momenta are rescaled by $M^2$. This was achieved in the recent paper~\cite{Marino:2025ido}. It is based on the expansion around the $O(N)$ broken vacuum~\cite{Jevicki:1977zn}, in a way similar to the traditional perturbative approach to the NLSM. It was shown that, after considering all the diagrams at NLO in $1/N$, one obtains the following IR finite correction to the self-energy:
\begin{align}
&\frac{1}{N}\langle \bPhi_0(p)\cdot \bPhi_0(-p) \rangle=S(p^2,M^2,\gamma) \ , \\
&S^{-1}(p^2,M^2,\gamma)=M^2\bigg(\mathfrak{p}^2+\frac{1}{N}\mathfrak{S}(\mathfrak{p}^2,\gamma)\bigg) \ ,
\end{align}
where we have introduced the dimensionless momentum $\mathfrak{p}$ through $p^2=M^2\mathfrak{p}^2$. The expression for the function $\mathfrak{S}(\mathfrak{p}^2,\gamma)$ can be found in~\cite{Marino:2025ido}. Here we quote the following form
\begin{align}
\mathfrak{S}(x,\gamma)=\frac{x}{x+1}-\bigg(\frac{2\ln(1+x)}{1+x}+\frac{x\ln x}{(x+1)^2}\bigg)\gamma +\sum_{n=1}^{\infty}\gamma^{n+1}(-1)^{n}\bigg(\frac{x^2\ln^{n+1}x}{(x+1)^{n+2}}+\tilde {\cal J}_n(x)\bigg) \ ,
\end{align}
where we have
\begin{align}
&\tilde {\cal J}_n(x)=\sum_{\ell \ge 1}{\rm Res}_{s=\ell} \bigg[(x)^s(\psi(-s)+\psi(s+1)+2\gamma_E){\cal M}[g_n(y);-s]\bigg] \ , \label{def:Jn}\\
& {\cal M}[g_n(y);-s]=\int_0^{\infty}{\rm d}y\frac{y^{-s} \ln ^n y}{(y+1)^{n+1}} \ .
\end{align}
Notice that the $\tilde {\cal J}_n(x)$ appearing here is almost the ${\cal J}_n(x)$ in~\cite{Marino:2025ido}, but with one factor of $\frac{(-1)^n x \ln^{n+1} x}{(x+1)^n}$ subtracted. We expect that the small momentum limit $\mathfrak{p}^2 \rightarrow 0$ of the perturbative self-energy above should reduce to the perturbative self-energy of NLSM, up to UV divergent pieces.
Let us show that this is the case. For this purpose, we need to expand the
functions $\tilde {\cal J}_n(x)$ at small $x$, up to first order in $x$. This is equivalent to picking up the residue at $s=1$ in Eq.~(\ref{def:Jn}). Due to the increasing powers of $\ln y$, the order of the poles at $s=1$ grows with $n$. After introducing the representations
\begin{align}
&\ln y^n=\frac{\rd^n}{\rd t^n} y^t \biggl|_{t=0} \ , \\
&\int_0^{\infty} {\rm d}y \frac{y^{-s}y^t}{(y+1)^{n+1}}=\frac{\Gamma (-s+t+1) \Gamma (n+s-t)}{\Gamma (n+1)}, \quad  -n+{\rm Re}(t)<{\rm Re}(s)<1+{\rm Re}(t) \ , \label{eq:MBJn}
\end{align}
the $s=1$ residues can be conveniently extracted as
\be
\ba
&{\rm Res}_{s=1} \bigg[(x)^s(\psi(-s)+\psi(s+1)+2\gamma_E){\cal M}[g_n(y);-s]\bigg] \nonumber \\
&=x\frac{\rd^n}{\rd t^n} \bigg[\frac{\Gamma (t) \Gamma (n-t+1)}{\Gamma (n+1)}-\frac{1}{t}\bigg]_{t=0}+x\frac{\rd^n}{\rd t^n}\bigg[x^t(-\psi(-1-t)-\psi(2+t)-2\gamma_E)+\frac{1}{t}\bigg]_{t=0} \ .
\ea
\ee
The first term is due to the pole at $s=1$ in Eq.~(\ref{eq:MBJn}), while the second term is due to the pole at $s=1+t$. Given the above, the small $x$ limit of $\mathfrak{S}(x,\gamma)$ up to order $x$ can be written as
\begin{align}
\mathfrak{S}(x,\gamma)\bigg|_{x\rightarrow 0}=x\mathfrak{S}^{0}(x,\gamma)+{\cal O}(x^2) \ ,
\end{align}
with
\begin{align}
&\mathfrak{S}^{0}(x,\gamma)=\sum_{n=1}^{\infty}(-1)^n \gamma^{n+1}\frac{\rd^n}{\rd t^n} \bigg[\frac{\Gamma (t) \Gamma (n-t+1)}{\Gamma (n+1)}-\frac{1}{t}\bigg]_{t=0} \nonumber \\
&+1+\sum_{n=0}^{\infty}(-1)^n \gamma^{n+1}\frac{\rd^n}{\rd t^n}\bigg[x^t(-\psi(-1-t)-\psi(2+t)-2\gamma_E)+\frac{1}{t}\bigg]_{t=0} \ .  \label{eq:quarticS}
\end{align}
We now show that the second line exactly corresponds to the self-energy of the NLSM. For this purpose, it is important
to notice that, at leading order in the $1/N$ expansion, one has
\begin{align}
\frac{1}{\gamma}=\ln \frac{M^2}{m_0^2}, \qquad   \frac{1}{\gamma}+\ln \mathfrak{p}^2=\ln \frac{p^2}{m_0^2}=\frac{2}{\lambda(p)}\ ,
\end{align}
which is just the running coupling constant in the NLSM with non-perturbative scale $m_0$. One then introduces the Borel transform for the second line in Eq.~(\ref{eq:quarticS})
\begin{align}
&\sum_{n=0}^{\infty}\frac{t^n}{n!}(-1)^n \frac{\rd^n}{\rd t^n}\bigg[x^t(-\psi(-1-t)-\psi(2+t)-2\gamma_E)+\frac{1}{t}\bigg]_{t=0}\nonumber \\
&=x^{-t}\bigg(\frac{1}{t-1}-\psi(1+t)-\psi(2-t)-2\gamma_E\bigg)+\frac{x^{-t}-1}{t} \ .
\end{align}
In this way, one can finally write, after changing $t=y/2$,
\begin{align}
\mathfrak{S}^{0}(x,\gamma)=Z(\gamma)-\ln \gamma+\gamma_E+\Phi_0(\lambda(p))\ ,
\end{align}
where $\Phi_0$ is exactly the two-point function of the NLSM, given in Eq.~(\ref{Phi0}), plus the ``UV-divergences'' given by
\begin{align} \label{eq:Z1}
Z(\gamma)=\sum_{n=1}^{\infty}(-1)^n \gamma^{n+1}\frac{\rd^n}{\rd t^n} \bigg[\frac{\Gamma (t) \Gamma (n-t+1)}{\Gamma (n+1)}-\frac{1}{t}\bigg]_{t=0}\ .
\end{align}
Notice that the infinite sum $Z(\gamma)$ is also divergent. This is very different from the RG
evolution factors in the $\overline{\rm MS}$ scheme, which are normally conjectured to be free from renormalons.

In fact, by including high-order terms in the small-$x$ expansion, which can be shown to have the following form
\begin{align}
\mathfrak{S}(x,\gamma)=x\sum_{k=0}^{\infty} x^k \bigg(Z_k(\gamma)+\frac{1}{\gamma^k}f_k(\lambda(p))\bigg) \ ,
\end{align}
all the singularities in the $Z_k$, including $Z_0=Z$, are cancelled:
\begin{align}
{\rm Res}\bigg[{\cal B}[Z_k](y)\re^{-\frac{y}{2\gamma}}\bigg]_{y=2n}+\frac{x^n}{\gamma^{n+k}}{\rm Res} \bigg[{\cal B}[f_{n+k}](y)\re^{-\frac{y}{\lambda(p)}}\bigg]_{y=2n}=0 \ .
\end{align}
On the other hand, the higher Borel singularities of ${\cal B}[f_{k}](y)$ at $y\ge 2(k+1)$ are not cancelled, and represent the genuine Borel singularities of $\mathfrak{S}(x,\gamma)$. Moreover, it is possible to show that they exactly sum to the renormalon ambiguity found in Eq.~(2.71) of~\cite{Marino:2025ido}.

Notice that if we change back to physical units and regard $a^2=M^{-2}$ as a ``lattice cutoff'',
with $\gamma=\gamma(a)$ as the bare coupling constant, then the small $x$ expansion above is in fact an expansion in $a^2p^2$:
\begin{align}
M^2\mathfrak{S}(x,\gamma)=p^2\sum_{k=0}^{\infty} (p^2a^2)^k \bigg(Z_k(\gamma(a))+\frac{1}{\gamma(a)^k}f_k(\lambda(p^2))\bigg) \ .
\end{align}
The continuum limit corresponds to the $k=0$ term, while all the terms with $k\ge 1$ can be regarded
as ``discretization effects''. The renormalons in $Z(\gamma)$ do not affect the continuum limit; they
rather measure the ambiguities for the $a^2p^2$ corrections.

\subsection{Non-perturbative corrections from the LSM}\label{sec:LSMcon}
Clearly, the matching of the perturbative corrections between the two models is the first
step to establish their relationship. We would like also to show that this relation holds at the non-perturbative level.
For this purpose one needs a non-perturbative formulation of the LSM as well.

This can be achieved by referring to the formulation of the quartic model in Eq.~(\ref{def:quartic}) with a positive squared
mass term. An important fact is that the small $\hat g$ expansion of this theory is Borel-summable~\cite{eckmann1975decay}, see~\cite{Serone:2018gjo,Serone:2019szm} for more
recent discussions. This allows to define the LSM as its strong coupling limit. In particular,
the two-point function in the $1/N$ expansion reads

\begin{align}
&S^{-1}_{\rm q}(p^2,m_0^2,\hat g)=p^2+m_0^2+\frac{1}{N}\Sigma^{1}_{\rm q}(p^2,m_0^2,\hat g)+ \CO(N^{-2}) \ , \\
&\Sigma^{1}_{\rm q}(p^2,m_0^2,\hat g)=\Sigma^{1}_{ {\rm q},S}(p^2,m_0^2,\hat g)+ \Sigma^{1}_{ {\rm q},T}(m_0^2,\hat g)
\end{align}
where
\be
\label{qST}
\ba
\Sigma^1_{ {\rm q},S}(p^2,m_0^2,\hat g)&=-\int \frac{\rd^2k}{(2\pi)^2}\frac{(4\pi m_0^2 \hat g)^2 F(k^2,m_0^2)}{1+4\pi \hat g m_0^2F(k^2,m_0^2)}\frac{1}{(p-k)^2+m_0^2}\ , \\
\Sigma^1_{ {\rm q},T} (m_0^2,\hat g)&=-\frac{2\pi \hat g m_0^2}{1+4\pi \hat gm_0^2 F(0,m_0^2)}\int \frac{\rd^2k}{(2\pi)^2}\frac{(4\pi m_0^2 \hat g)^2 F(k^2,m_0^2)}{1+4\pi \hat g m_0^2F(k^2,m_0^2)}\frac{\partial}{\partial m_0^2}F(k^2,m_0^2)
\ea
\ee
are the contributions from the self-energy and tadpole diagrams, respectively, and
the function $F(k^2,m_0^2)$ is the massive version of the bubble:
\begin{align}
F(k^2,m_0^2)=\frac{1}{2}\int \frac{\rd^2l}{(2\pi)^2}\frac{1}{(l^2+m_0^2)((k-l)^2+m_0^2)} \ .
\end{align}
The subscript ${\rm q}$ in the above equations indicates that they refer to the quartic LSM. In~\cite{Marino:2025ido}, by
using this formulation, it was shown that at NLO in the $1/N$ expansion, the physical
mass of the theory remains of order $m_0^2$, up to logarithmic corrections. In particular, all the power divergences cancel:
\begin{align}\label{eq:massLSM}
m^2=M^2\re^{-\frac{1}{\gamma}}+\frac{M^2\re^{-\frac{1}{\gamma}}}{N}\bigg(-2Z(\gamma)-2\ln \gamma-\frac{2}{\gamma}+4\ln 2+2\gamma_E\bigg)+{\cal O}\left(\frac{1}{N^2}\right) \ .
\end{align}
Comparing this with the mass relation~(\ref{eq:SMmass1}) of the NLSM in the SM scheme, we
find perfect agreement for the coefficients of the terms $\ln \gamma$ and $1/\gamma$, which are
due to the universal $1/N$ corrections to the one-loop and two-loop beta function. The
remaining divergences in $Z(\gamma)$ allow to define the higher order corrections to the
beta function in a LSM regularization scheme for the NLSM.

Moreover, one can go beyond the mass correction, and by using the Mellin-Barnes (MB)
representations in~\cite{Liu:2025edu}, we
can show that the non-perturbative corrections to the LSM two-point function also reduce to the NLSM version, in the
$\gamma \rightarrow 0$ limit. In particular, we found that, up to the first exponentially small correction, one has
\begin{align}\label{eq:limitLSM}
&S^{-1}(p^2,M^2,\gamma)\bigg|_{\gamma \rightarrow 0}\nonumber \\
&=(p^2+m^2)+\frac{1}{N} \bigg(Z(\gamma)-\ln \gamma+\gamma_E\bigg)(p^2+m^2)+\frac{m^2}{N}S\left(\frac{p^2}{m^2}\right)+{\cal O}(a^2p^2) \ ,
\end{align}
where $S(p^2/m^2)$ is exactly the $1/N$ correction to the self-energy of NLSM in the SM scheme
given in Eq.~(\ref{Sx}). In particular, we can reproduce both the mass correction
in Eq.~(\ref{eq:massLSM}), and the trans-series $\Phi_1(\lambda)$ in Eq.~(\ref{Phi1}) for $S$. Moreover, the UV divergences factorize to the factor $Z(\gamma)-\ln \gamma$, in agreement with the multiplicative renormalizability. The relation Eq.~(\ref{eq:limitLSM}) allows the inclusion of the $a^2$ corrections as well, which will cancel the quadratic renormalons in $Z(\gamma)$.

To summarize, in the $\gamma \rightarrow 0$ limit, the LSM indeed becomes the NLSM at the non-perturbative mass scale $m^2$, and UV fluctuations decouple into multiplicative renormalization factors up to leading order in the small-$a^2$ expansion. However, there are renormalons in these factors, suggesting intrinsic ambiguities in defining the ${\cal O}(a^2)$ discretization effects, in a way similar to the ambiguities in defining the condensates. We expect this to be a general feature of cutoff regularization schemes in asymptotically free QFTs.

\subsection{Power divergences and the UV renormalon from the LSM}\label{sec:renorUV}

The LSM regularization of NLSM is a cutoff regularization. As such, there could be power divergences in
the computation. The appearance of these divergences confirms the nature of the UV renormalon
of the perturbative self-energy at $y=2$, and the picture for its cancellation developed in Sec. \ref{sec:UVrenor}.

Indeed, if we consider the self-energy and tadpole contributions (\ref{qST}) separately,
then in the $\hat g \rightarrow \infty$ or $\gamma \rightarrow 0$ limit, both of them contain
power divergences. These can be extracted by using the MB method, and one finds:
\begin{align}
&\Sigma^1_{ {\rm q}, S}(p^2,m_0^2,\hat g)=-M^2\gamma\bigg(\frac{1}{\gamma}+Z(\gamma)\bigg)+(Z(\gamma)-\ln \gamma+\gamma_E)(p^2+m_0^2)+\left(p^2\Phi_0(\lambda)+...\right) \ , \\
&m_0^2+\frac{1}{N}\Sigma^1_{ {\rm q}, T}(m_0^2,\hat g)=m^2+\frac{1}{N}M^2\gamma\bigg(\frac{1}{\gamma}+Z(\gamma)\bigg)\mp \frac{1}{N}\ri\pi m^2 \ , \label{eq:tadLSM}
\end{align}
where we have kept only the power divergences and finite contributions,
but neglected all the $a^2$ corrections. In Appendix~\ref{sec:MB} we show
how to extract the power divergent part of the tadpole diagram contribution
$\Sigma_T$ by using the MB method. The MB representation provides the following form for $Z(\gamma)$,
\begin{align}\label{eq:Z2}
Z(\gamma)=\int_{0}^{\infty}  \re^{-\frac{t}{\gamma}}\bigg(\left(\frac{t}{\gamma}\right)^t\Gamma(-t)+\frac{1}{t}\bigg) \rd t \ ,
\end{align}
which can be shown to be equivalent to Eq.~(\ref{eq:Z1}). By using this
representation, one can see that the $y=2$ singularity of $\Phi_0$ is exactly cancelled by the
$t=1$ singularity of $-M^2\gamma Z(\gamma)$, if we use in addition that
$m_0^2=M^2 \re^{-1/\gamma}$ at this order in the $1/N$ expansion.
Similarly, the ambiguity of the tadpole diagram, given by $\mp \ri\pi m_0^2$, which is
exactly the ambiguity of the $X$ condensate, is also cancelled by the power divergence
of the tadpole diagram. The fact that the $y=2$ ambiguity of the perturbative self-energy
cancels with power divergences implies that this is an UV renormalon, instead of an
IR renormalon. This confirms the conclusion of the discussion in Sec. \ref{sec:UVrenor}:
the cancellation with the $X$ condensate is therefore a cancellation
between two UV renormalons, and it is a consequence of the multiplicative renormalizability of the theory.

\section{Conclusion} \label{sec:concl}
In this paper we have shown how, in the two-dimensional NLSM, the large momentum
expansion obtained from the exact large $N$ non-perturbative solution can be reproduced from a
massless OPE computation with condensates. This is achieved through the representation of the NLSM as a limit of
the quartic LSM at the integrand level. One crucial point of the method is that one has to sum over infinitely
many scale-less condensates $\langle (\bPhi^2)^k \rangle$, and this sum is then transmuted into a coupling constant
expansion through the constraint Eq.~(\ref{constraint}). As such, the OPE computation can be performed in a
manifestly $O(N)$ symmetric manner, without solving the constraint equation explicitly.
This makes it easier to obtain the non-perturbative corrections due to condensates. On the other hand, at each exponentially small order in the OPE, the number of scale-full operators multiplying the scaleless operators $(\bPhi^2)^k$ can still be kept finite.
In particular, at the first exponentially small order, the only such scale-full operator is the Lagrangian operator
$X$ (or equivalently, the trace anomaly operator), and the non-trivial perturbative series $\Phi_1(\lambda)$ is entirely
due to perturbative corrections to its coefficient function.
The situation is similar to the OPE of the current-current correlator in massless QCD, in which the only condensate that contributes at the next-to-leading order in the OPE is exactly the Lagrangian operator $F^2$. As in the case of the Gross-Neveu model~\cite{tsc}, by comparing the OPE computation with the trans-series obtained from the exact large $N$ method, one can determine the condensate of the Lagrangian operator and show consistency with the free energy computation of \cite{biscari}.

On the other hand, and in contrast to the QCD example, the leading $y=2$ renormalon of the
perturbative self-energy of the NLSM is an UV renormalon, despite the fact that it cancels the
ambiguity of the $X$ condensate. This looks surprising, because UV renormalons
usually reflect ambiguities in defining the continuum limit of a theory with an UV cutoff, while the two-point
function of the NLSM is clearly well defined and free from such ambiguities. The representation of the
NLSM as a limit of the LSM also helps to resolve this puzzle: if we first perform the loop integrals, then the ``Higgs mass'' of the quartic LSM becomes an UV cutoff for the NLSM, and one can keep track of the UV renormalons from the power divergences. Indeed, we found power divergences in the self-energy and the tadpole diagrams separately, which cancel respectively with the $y=2$ renormalon
of the perturbative self-energy and the ambiguity in the $X$ condensate. This implies that both, the renormalon at $y=2$ and the ambiguity,
have to be regarded as UV renormalons. However, a non-trivial UV property of the NLSM, the multiplicative renormalizability, requires the cancellation of power divergences between the two diagrams. This then leads to the cancellation of the two UV renormalons between condensates and coefficient functions. To maintain the correct UV behavior, a perturbative and a non-perturbative contribution have to ``match,'' not across the gap between IR and UV scales, but at a common UV point much higher than both. This fact suggests that correlations between UV and IR contributions in the OPE, or in more general power expansions, could happen more often and be more complicated than expected.

Our work could be generalized in various ways. The representation of the NLSM as a limit
of LSM allows to perform the OPE computation for the NLSM in an $O(N)$ symmetric manner and
makes the inclusion of operator condensates simpler. However, it is possible to obtain the correct perturbative contribution to the self-energy (\ref{eq:NLSMpert}) by working in a more conventional perturbative framework and expanding around the $O(N)$ broken vacuum \cite{mmr-unpublished}. It would be interesting to see how the operator condensates can also be incorporated into this perturbative approach.

Similarly, the weakly coupled LSM at the Higgs mass scale also allows perturbative expansions in the coupling constant $\gamma$. In~\cite{Jevicki:1977zn,Marino:2019fvu,Marino:2025ido}, a perturbative framework based on expanding around the $O(N)$ broken vacuum was developed to compute such expansions. It would also be interesting to see if the perturbative series for the LSM at the Higgs mass scale can also be computed in an $O(N)$-symmetric perturbative framework.

The LSM model, as a quartic model, has its own super-renormalizable OPE limit. It was noticed
recently~\cite{Liu:2025edu} that this limit also suffers from a large number of logarithms that generate factorial enhancements in the OPE.  These logarithms in the quartic model are quite similar to the logarithms in the NLSM: in the massive formulations, they are both due to the intermediate momentum region of the massive bubbles in the large $p^2$ limit, while in the massless OPE computations they are all generated by the IR divergences of the massless bubbles and are subtracted by using the same $\langle (\bPhi^2)^k \rangle$ condensates. There seems to be a logarithmic renormalization group flow in the LSM, from the quartic model OPE limit in the UV, to the NSLM OPE limit in the IR. A considerable amount of logarithms survive, but simultaneously reorganize along the flow. It would be interesting to better understand such an RG flow.

We should also notice that the LSM regularization leads to multiplicative logarithmic
divergences with IR renormalon ambiguities. This is very different from the renormalization data in the
$\overline{\rm MS}$ scheme, which are usually conjectured to be free from renormalons. Such ambiguities do not affect the
continuum limit itself, but correlate with ambiguities in the ${\cal O}(a^2)$ ``discretization effects''. We expect this to be a general feature of cutoff regularizations, including the lattice regularization. In the lattice QCD literature, renormalons in the power divergences for operator condensates and in the Wilson-line self-energy are well known (see~\cite{Bauer:2011ws,Bali:2014fea} for strong numerical evidence), but are rarely discussed in logarithmic divergences. As such, this point requires further investigation. For this purpose, it would be interesting to study the continuum limit of the the NLSM with a lattice regularization, at NLO in the $1/N$ expansion, as in \cite{biscari, cr-review}, and
explicitly work out the power divergences and the renormalons in the logarithmic divergences.

Finally, in the NLSM itself, our work is clearly limited to the NLO in the $1/N$ expansion. Our formalism could be generalized to higher orders, by including the scale-less condensates. Since to any given order in the perturbative expansion only a finite numbers of orders in the $1/N$ expansion are required, our formalism could be used in principle to compute the traditional perturbative series at higher loops, in an $O(N)$-symmetric manner. On the other hand, extending the OPE calculation beyond the NLO in the $1/N$ expansion is difficult. One might expect that higher order contributions could be obtained more efficiently by considering the form factor expansion in the large $p^2$ limit directly, since to each order in the $1/N$ expansion only a finite number of intermediate state particles are involved. However, studying the OPE and extracting exponentially small contributions from form factors is still a challenging task (see e.g. \cite{balog-weisz} for results along this direction). The case $N=3$ deserves special attention. In this case, one expects to have instanton contributions, as shown recently in the case of the free energy ~\cite{Marino:2022ykm,bbv3}. It remains to be seen how these contributions would be incorporated in the OPE and if there are simple frameworks allowing the computation of such effects.

\section*{Acknowledgements}
We would like to thank Martin Beneke, Ramon Miravitllas, Tom\'as Reis and Marco Serone for very useful discussions
and for their comments on a preliminary version of this paper. M.M. is particularly grateful to Ramon Miravitllas
and Tom\'as Reis for collaborating in an
earlier attempt \cite{mmr-unpublished} to
solve the problem considered in this paper. We would also like to thank the anonymous referees for their comments and suggestions, which helped us very much to improve the presentation. The work of M.M. has been supported in part by the ERC-SyG project
``Recursive and Exact New Quantum Theory" (ReNewQuantum), which
received funding from the European Research Council (ERC) under the European
Union's Horizon 2020 research and innovation program,
grant agreement No. 810573.

\appendix
\sectiono{Sums over bubbles}
\label{pmp-trick}

In \cite{pmp, pm}, a useful technique was introduced to obtain the renormalized
perturbative series associated to a chain of bubbles. Here we summarize the main results.

Let us consider a sum of corrections in the bare 't Hooft coupling $\lambda_0$, of order $1/N$:
\be
\sum_{n \ge n_0} a_n (p^2)   \lambda_0^n \ ,
\ee
where $n_0 \ge 1$. The choice of $\lambda_0$ is such that, at leading order in the $1/N$ expansion, the renormalization function
is given by
\be
\label{zl}
Z^{-1}_\lambda= 1+ {\lambda \over \epsilon} + \CO(N^{-1}) \ .
\ee
Then, the renormalized sum is of the form
\be
\label{ren-sum}
\sum_{n \ge n_0} (\nu^2)^{n\epsilon/2} a_n (p^2)  Z_\lambda^n \lambda^n \ .
\ee
Let us assume that one can find a ``structure function" $F(x,y)$, which is analytic in both arguments at $x=0$, $y=0$, and such that one
can write
\begin{equation}
(\nu^2)^{n\epsilon/2} a_n (p^2) = \frac{F(\epsilon,n\epsilon)}{n\epsilon^n} \ .
\label{eq_structure_function}
\end{equation}
We also define
\be
F(x,y)= \sum_{j \ge 0} F_j(x) y^j \ .
\ee
Then, one can show that
\be
\label{main-f}
\ba
&\sum_{n \ge n_0} {F(\epsilon, n \epsilon) \over n \epsilon^n} \lambda^n \left(1+{ \lambda \over\epsilon} \right)^{-n}\\
&= F_0(\epsilon)\ln\left(1+{\lambda \over \epsilon} \right)-F_0(\epsilon) \sum_{m=1}^{n_0-1}{(-1)^{m-1}  \over m}  \left( {\lambda \over \epsilon} \right)^m   + \sum_{m \ge n_0} (m-1)! F_m (\epsilon) \lambda^m+ \CO(\epsilon) \ .
\ea
\ee
The first two terms in the second line of (\ref{main-f}) lead to divergences as $\epsilon \to 0$. We note the useful formula
\be
\label{F0div}
\left[ F_0(\epsilon) \ln\left(1+ {\lambda \over \epsilon} \right) \right]_{\rm div}=  \int_0^\lambda {F_0(-u) \over u+ \epsilon} \rd u \ ,
\ee
and we can write the divergent part in (\ref{main-f}) as
\be
\label{div-part}
  \int_0^\lambda {F_0(-u) \over u+ \epsilon} \rd u - \sum_{m=1}^{n_0-1} {(-1)^{m-1} \over m} \lambda^m \sum_{k=0}^{m-1} F_{0,k} \epsilon^{k-m}\ ,
  \ee
where the coefficients $F_{0,k}$ are defined by
\be
F_0(\epsilon)=  \sum_{k \ge 0} F_{0,k} \epsilon^k.
\ee
The finite part of the second line in (\ref{main-f}) can be written as
\be
\label{finite-part}
-\int_0^\lambda {F_0(-u) -F_{0,0} \over u} \rd u- \sum_{m=1}^{ n_0-1} {(-1)^{m-1} \over m} F_{0,m} \lambda^m+ \sum_{m \ge n_0} (m-1)! F_m (0) \lambda^m \ .
\ee
We note that the last sum has the form of an inverse Borel transform:
\be
\ba
\sum_{m \ge n_0} (m-1)! F_m (0) \lambda^m&= \int_0^\infty \re^{-y/\lambda} \left( \sum_{m \ge n_0} F_m (0) y^{m-1} \right) \rd y \\
&=
\int_0^\infty \re^{-y/\lambda} \left( F(y)- \sum_{m =0}^{n_0-1} F_m (0) y^{m}  \over y \right) \rd y \ ,
\ea
\ee
where
\be
F(y)= F(0,y) \ .
\ee

\section{MB analysis of the tadpole diagram in the LSM} \label{sec:MB}

In this appendix we show how to extract the power divergences of the tadpole contribution Eq.~(\ref{eq:tadLSM}),  by using the MB method. The tadpole contribution to the self-energy in the quartic LSM reads
\begin{align}
\Sigma_{{\rm q}, T}^{1}(m_0^2,\hat g)=\frac{1}{2}\frac{g}{1+gF(0)}\int \frac{\rd^2k\rd^2l}{(2\pi)^4}\frac{1}{(l^2+m_0^2)^2 \left((k-l)^2+m_0^2\right)}\frac{g^2F(k^2)}{1+gF(k^2)} \ ,
\end{align}
 where we have introduced the dimensionful coupling
\begin{align}
g=4\pi m_0^2\hat g \ .
\end{align}
Notice that the term $g^2F$ in the numerator is due to the normal-ordering prescription for the mass renormalization in the quartic model. The relation
\begin{align}
\int \frac{\rd^2l}{(2\pi)^2}\frac{1}{(l^2+m_0^2)^2\left((k-l)^2+m_0^2\right)}=\frac{2F(k^2)}{k^2+4m_0^2}+\frac{1}{4\pi} \frac{1}{(k^2+4m_0^2) m_0^2}
\end{align}
is also useful. Then, by using $F(0)=(8\pi m_0^2)^{-1}$ and
\begin{align}
\int \frac{\rd^2k}{(2\pi)^2}\frac{F(k^2)}{k^2+4m_0^2}=\frac{1}{(4\pi)^2 m_0^2}\ln 2 \ ,
\end{align}
one can write
\begin{align}
&\Sigma_{{\rm q}, T}^{1}=m_0^2\frac{\hat g^2\ln 4}{\hat g+2}+\frac{\hat g-2}{\hat g+2} {\cal I}(m_0^2,\hat g) \ , \\
& {\cal I}(m_0^2,\hat g)=\int \frac{\rd^2k}{(2\pi)^2}\frac{g^2F(k^2)}{1+gF(k^2)}\frac{1}{k^2+4m_0^2} \ .
\end{align}
The integral ${\cal I}$ can be expanded by using the MB representation of $\frac{g^2F(k^2)}{1+gF(k^2)}$~\cite{Liu:2025edu}. One has
\begin{align}
&{\cal I}(m_0^2,\hat g)=-m_0^2\int \frac{\rd s_1}{2\pi i }\int_0^{\infty} \frac{\rd t}{s_1}t^{-s_1}\frac{\rd}{\rd t}\int\frac{\rd s}{2\pi i}\bigg(\frac{\hat g^{s_1+1}\pi  4^{s-s_1}}{\sin \pi  (s-s_1)}{\cal M}(s,s_1,t)\bigg) \ , \\
&{\cal M}(s,s_1,t)=\frac{\pi \Gamma (s_1-2 s) \Gamma (s+t) \, _2\tilde{F}_1(s_1-1,s+t;-s+s_1+t;-1)}{\sin (\pi s_1) \Gamma (-s_1)} \ .
\end{align}
The integration contour for the MB variables should be chosen as $0<2{\rm Re}(s)<{\rm Re}(s_1)<1$.  From the above, one can obtain the expansion in the limit $\hat g\rightarrow \infty$ by shifting the contour of the $s_1$ variable to the left. Notice that, when shifting $s_1 \le -\frac{1}{2}$, the small-$t$ behavior becomes regular enough and the $\frac{\rd}{\rd t}$ derivative is not needed. However, for the poles at $s_1=0$; $s_1=s$ (followed by $s=-t$, $s=0$); and $s_1=2s$ (followed by $s=0$), the small-$t$ behavior is still rather singular and one needs the derivatives. These poles lead precisely to the power divergences. After combining them together, one obtains the following Borel integrand for the power divergences
\begin{align}
{\cal I}_{-1}(t)=m_0^2\hat g\bigg(g_1(t)+g_2(t)+\sum_{k=0}^{\infty}f_k(t)\bigg) \ ,
\end{align}
with
\begin{align}
& g_1(t)={\rm Res}_{s=-t}t^{-s}\frac{\rd}{\rd t}\bigg(\frac{\pi  \left(\hat g\right)^{s}t \Gamma \left(\frac{s+t}{2}\right)}{2 s\sin \pi s\Gamma \left(\frac{1}{2} (-s+t+2)\right) }\bigg)\ , \\
& g_2(t)={\rm Res}_{s=0}t^{-s}\frac{\rd}{\rd t}\bigg(\frac{\pi  \left(\hat g\right)^{s}t \Gamma \left(\frac{s+t}{2}\right)}{2 s\sin \pi s\Gamma \left(\frac{1}{2} (-s+t+2)\right)}+\frac{\pi  \left(\hat g\right)^{s}2 \Gamma \left(\frac{1}{2} (s+t+1)\right)}{2 s\sin \pi s\Gamma \left(\frac{1}{2} (-s+t+1)\right) }\bigg) \ , \\
&f_k(t)=\frac{\rd}{\rd t} \bigg(\frac{\pi  2^{2 s+1} t \Gamma (-2 s) \Gamma (s+t)}{\sin \pi s \Gamma (-s+t+1)}\bigg)_{{\rm Res}_{s=-t-k}+{\rm Res}_{s=-k}} \ .
\end{align}
Notice that the $f_k$ contribution is due to the poles at $s_1=0$ followed by $s=-k$ and $s=-t-k$, and for the first two terms we have expressed the $_2F_1(s-1,s+t;t;-1)$ in terms of gamma functions. In fact, the hypergeometric functions $_2F_1(s_1-1,s+t;t+s_1-s;-1)$ can all be expressed in terms of gamma functions for $s_1\in \mathbb{Z}_{\le 0}$ as well as $s_1 \in  s-\mathbb{Z}_{\ge 0}$. We first consider the $g_1+g_2+f_0$ combination. In this combination, both the large $t$ and small $t$ singularities are cancelled. In particular, we have
\begin{align}
g_2(t)+f_0(t)=\frac{\pi 4^{-t}}{\sin \pi t }(\pi  \cot (\pi  t)+\ln (4))-\frac{1}{t^2}+\frac{1}{t^2} \ , \\
\int_0^{\infty} \rd t \bigg( \frac{\pi 4^{-t}}{\sin \pi t }(\pi  \cot (\pi  t)+\ln (4))-\frac{1}{t^2}\bigg)=-\ln 4 \ .
\end{align}
On the other hand, the term $1/t^2$ combines with $g_1(t)$ to produce
\begin{align}
\int_0^{\infty} \left(\frac{1}{t^2}+g_1(t)\right) \rd t =\frac{1}{v}-\int_0^{\infty} \rd t \left(\frac{v\pi  \re^{-t+t v \ln(t)} }{\sin (\pi v t ) \Gamma (1+vt )}-\frac{\re^{-t}}{t}\right)\ .
\end{align}
Here we introduced the MB coupling constant $v$, defined by
\begin{align}
\frac{1}{v}+\ln v=\ln \hat g \ .
\end{align}
It agrees with $\gamma$ up to exponentially small and $1/N$ corrections. Furthermore, when $k \ge 1$, one can show that the integrands $f_k(t)$ are total derivatives of functions which decay fast at large $t$ and vanish at $t=0$, thus these contributions all vanish after integration. For example, for $k=1$ one has
\begin{align}
f_1(t)=\frac{\rd}{\rd t}\bigg(\frac{1}{2-2 t^2}-\frac{\pi 4^{-t} t}{2\sin \pi t }\bigg)\ ,
\end{align}
and it integrates to $0$. As such, the power divergences of ${\cal I}$ are entirely due to the combination $g_1+g_2+f_0$ and read
\begin{align}
{\cal I}_{-1}(m_0^2,\hat g)=m_0^2\hat g \bigg(-\ln4 +\frac{1}{v}+Z(v)\bigg).
\end{align}
The $-\ln 4$ cancels the first term in $\Sigma_{{\rm q}, T}^{1}$, which leads to the total power divergences
\begin{align}
\Sigma_{{\rm q}, T;-1}^1(m_0^2,\hat g)=m_0^2 \hat g \bigg(\frac{1}{v}+Z(v)\bigg)\ .
\end{align}
This agrees with Eq.~(\ref{eq:tadLSM}) after noticing $m_0^2 \hat g=M^2 \gamma$.

\bibliographystyle{JHEP}

\linespread{0.6}
\bibliography{biblio-qft}

\end{document}